\documentclass[epj]{svjour}
\usepackage{graphicx}
\usepackage{amssymb}
\begin{document}
\title{Aging, rejuvenation, and memory effects in short-range Ising spin
glass: Cu$_{0.5}$Co$_{0.5}$Cl$_{2}$-FeCl$_{3}$ graphite
bi-intercalation compound}
\author{M. Suzuki \and I. S. Suzuki
}                     
%
%
\institute{Department of Physics, State University of New York at
Binghamton, Binghamton, New York 13902-6000 USA,
\email{suzuki@binghamton.edu}}
\date{Received: date / Revised version: date}
%
\abstract{Non-equilibrium aging dynamics in 3D Ising spin glass
Cu$_{0.5}$Co$_{0.5}$Cl$_{2}$-FeCl$_{3}$ GBIC has been studied by zero-field
cooled (ZFC) magnetization and low frequency AC magnetic susceptibility ($f
= 0.05$ Hz), where $T_{g} = 3.92 \pm 0.11$ K. The time dependence of the
relaxation rate $S(t) = (1/H)$d$M_{ZFC}/$d$\ln t$ for the ZFC magnetization
after the ZFC aging protocol, shows a peak at a characteristic time
$t_{cr}$ near a wait time $t_{w}$ (aging behavior), corresponding to a
crossover from quasi equilibrium dynamics to non-equilibrium.  The time
$t_{cr}$ strongly depends on $t_{w}$, temperature ($T$), magnetic field
($H$), and the temperature shift ($\Delta T$).  The rejuvenation effect is
observed in both $\chi^{\prime}$ and $\chi^{\prime\prime}$ under the
$T$-shift and $H$-shift procedures.  The memory of the specific spin
configurations imprinted during the ZFC aging protocol can be recalled when
the system is re-heated at a constant heating rate.  The aging,
rejuvenation, and memory effects observed in the present system are
discussed in terms of the scaling concepts derived from numerical studies
on 3D Edwards-Anderson spin glass model.
} 
\titlerunning{Aging, rejuvenation, and memory effects in 
Cu$_{0.5}$Co$_{0.5}$Cl$_{2}$-FeCl$_{3}$ GBIC}
\maketitle
\section{\label{intro}Introduction}
In recent years, non-equilibrium dynamics, in particular, aging dynamics,
of spin glass (SG) systems has been extensively studied theoretically
\cite{Fisher1988,Komori1999,Yoshino2001,Picco2001,Bernardi2001,Yoshino2002,Takayama2002,Takayama2003,Berthier2003}
and experimentally 
\cite{Lundgren1990,Hammann1990,Nordblad1986,Granberg1988,Djurberg1999,Joh1999,Jonsson1999,Jonason2000,Bouchaud2001,Dupuis2001,Jonsson2002a,Jonsson2002b,Zotev2002,Zotev2003,Bert2003,Rodriguez2003,Suzuki2003,Kenning2004,Jonsson2003}. 
When the SG system is quenched
from a high temperature above the SG transition temperature $T_{g}$ to a
low temperature $T$ below $T_{g}$ (this process is called the zero-field
cooled (ZFC) aging protocol), the initial state is not thermodynamically
stable and relaxes to more stable state.  The aging behaviors depend
strongly on their thermal history within the SG phase.  A typical
experimental method for the study of the aging dynamics is the time
dependence of the ZFC magnetization ($M_{ZFC}$) after the ZFC aging
protocol.  When a small magnetic field ($H$) is applied after an isothermal
aging at a temperature $T$ below $T_{g}$ for a wait time $t_{w}$, $M_{ZFC}$
increases as an observed time ($t$) increases.  The rejuvenation (chaos)
and memory effects are also significant features of the aging dynamics. 
These effects are typically measured from the low frequency AC magnetic
susceptibility.

Such an aging dynamics of SG phase is explained mainly in terms of a
real-space picture (the droplet model) \cite{Fisher1988}.  In this
picture, the SG coherence length for equilibrium SG order, grows up slowly
in aging processes.  The scaling properties of age-dependent macroscopic
susceptibility can be described by a growing coherence length $L_{T}(t)$. 
The droplet model also predicts the following two.  (i) The equilibrium SG
states at two temperatures with the difference $\Delta T$ are uncorrelated
in length scales larger than the overlap length $L_{\Delta T}$, i.e.,
so-called temperature ($T$)-chaos nature of the SG phase.  (ii) In the
equilibrium and thermodynamic limits the SG phase is broken by a static
magnetic field $H$ of infinitesimal strength, thereby introduced is the
crossover length $L_{H}$.  This length separates the mean size of SG
domains $L_{T}(t)$, such that they are dominated by the Zeemann energy for
$L_{T}(t) > L_{H}$ and by the SG free energy gap for $L_{T}(t) \ll L_{H}$. 
Recently Monte Carlo (MC) simulations \cite{Takayama2002,Takayama2003} on
the $T$- and $H$-shift aging processes have been carried out for the
three-dimensional (3D) Ising Edwards-Anderson (EA) SG model with Gaussian
nearest-neighbor interactions with zero mean and variance $J$, where $J$ is
in the unit of energy.  In the $T$-shift process, only a precursor of the
temperature-chaos effect is observed, while the results on the $H$-shift
process strongly support the droplet picture that the SG phase under a
finite $H$ is unstable in the equilibrium and thermodynamic limits.

In this paper we report our results on the aging dynamics (aging,
rejuvenation and memory) of a 3D Ising SG,
Cu$_{0.5}$Co$_{0.5}$Cl$_{2}$-FeCl$_{3}$ graphite bi-intercalation compound
(GBIC).  The aging behavior of the ZFC magnetization has been measured at
various aging processes including the $T$- and $H$-shift
perturbations.  The rejuvenation effect of the low frequency AC magnetic
susceptibility has been measured at aging processes including the $T$-
and $H$-shift perturbations.  The memory of the specific spin
configurations imprinted during the ZFC aging protocol can be recalled when
the system is re-heated at a constant heating rate.  The existence of the
characteristic lengths $L_{\Delta T}$ and $L_{H}$ is examined by the
$T$- and $H$-shift aging protocols.  The relaxation rate defined by
$S(t) = (1/H)$d$M_{ZFC}/$d$\ln t$ (see Sec.~\ref{backB} for more
definitions) exhibits a peak at a characteristic time $t_{cr}$, which is
comparable to a wait time $t_{w}$.  We show that the ratio $t_{cr}/t_{w}$
strongly depends on the aging process.  These results are compared with
those reported for typical spin glass systems such as Cu (1.5-13.5 at. \%
Mn)
\cite{Nordblad1986,Granberg1988,Djurberg1999,Joh1999,Zotev2002,Zotev2003,Rodriguez2003,Kenning2004},
Ag (11 at.  \% Mn) \cite{Jonsson1999,Jonsson2002b,Jonsson2003},
CdCr$_{1.7}$In$_{0.3}$S$_{4}$ \cite{Jonason2000,Bouchaud2001},
Fe$_{0.5}$Mn$_{0.5}$TiO$_{3}$ \cite{Bernardi2001,Dupuis2001,Jonsson2002a},
as well as results from MC simulations
\cite{Komori1999,Picco2001,Takayama2002,Takayama2003}.

The equilibrium dynamics of Cu$_{0.5}$Co$_{0.5}$Cl$_{2}$-FeCl$_{3}$ GBIC
has been reported in a previous paper \cite{Suzuki2003}.  The aging
dynamics has been also studied: the $\omega t$-scaling of $\chi^{\prime}$
and $\chi^{\prime\prime}$ is also confirmed, where $\omega$ is the angular
frequency of the AC magnetic field.  This compound undergoes a SG
transition at $T_{g}$ ($= 3.92 \pm 0.11$ K).  The system shows a dynamic
behavior that has some similarities and some significant differences
compared to a 3D Ising SG. It shows critical slowing down with a value of
the dynamic critical exponent that is rather similar to an Ising SG: $z =
6.6 \pm 1.2$, $\theta = 0.13 \pm 0.02$, and $\psi = 0.24 \pm 0.02$, where
$z$, $\psi$, and $\theta$ are the dynamic critical exponent, the stiffness
exponent (energy exponent) which determines the free energy of a droplet
excitation, and the barrier exponent which describes how the barrier
heights change with length scale.  The critical relaxation time $\tau$ is
well described by $\tau = \tau^{*}(1-T/T_{g})^{-x}$ with the dynamic
critical exponent $x = 10.3 \pm 0.7$ and $\tau^{*} = (5.29 \pm 0.07) \times
10^{-6}$ sec.  The in-field dynamics indicates, as for an
Ising SG, that the SG transition is destroyed by a magnetic field.  The
equilibrium dynamics shows a frequency dependence that is different from an
Ising SG, the absorption decreases with increasing frequency, whereas
ordinary 3D Ising SG shows a increasing absorption with increasing
frequency.  An aging behavior is observed that is rejuvenated by a large
enough (magnetic field) perturbation.

\section{\label{back}Background}
\subsection{\label{backA}Scaling Properties}
We present a simple review on the aging behavior of SG phase after the ZFC
aging protocol, based on the droplet model
\cite{Fisher1988,Komori1999,Lundgren1990}.  This ZFC aging protocol process
to the SG phase is completed at $t_{a}$ = 0, where $t_{a}$ is defined as an
age (the total time after the ZFC aging protocol process).  Then the system
is aged at $T$ under $H$ = 0 until $t_{a}$ = $t_{w}$, where $t_{w}$ is a
wait time.  Correspondingly, the size of domain defined by $R_{T}(t_{a})$
grows with the age of $t_{a}$ and reaches $R_{T}(t_{w})$ just before the
field is turned on at $t= 0$ or $t_{a} = t_{w}$.  The aging behavior in
$M_{ZFC}$ is observed as a function of the observation time $t$.  After $t
= 0$, a probing length $L_{T}(t)$ corresponding to the maximum size of
excitation grows with $t$, in a similar way as $R_{T}(t_{a})$.  When
$L_{T}(t) \ll R_{T}(t_{w})$, quasi-equilibrium dynamics is probed, but when
$L_{T}(t) \gg R_{T}(t_{w})$, non-equilibrium dynamics is probed.  It is
theoretically predicted that the mean SG domain-size $L_{T}(T)$ is
described by a power law given by \cite{Komori1999}
\begin{equation} 
L_{T}(t)/L_{0} \approx (t/t_{0})^{1/z(T)},
\label{eq:one} 
\end{equation} 
or by \cite{Fisher1988}
\begin{equation} 
L_{T}(t)/L_{0}\approx [(T/\Delta_{g})\ln(t/t_{0})]^{1/\psi},
\label{eq:two} 
\end{equation} 
where $L_{0}$ and $t_{0}$ are microscopic length and time scale, the
exponent $1/z(T)$ ($=bT/T_{g}$ with $b = 0.16$) \cite{Komori1999} linearly
depends on $T$ except for the region near $T_{g}$, and $\Delta_{g}$ and
$\psi$ are the characteristic scale of energy barrier of droplet
excitations and the associated exponent, respectively.  It is predicted by
Komori et al.  \cite{Komori1999} that $\Delta\chi^{\prime\prime}$ obeys the
$\omega t$-scaling form given by $\Delta\chi^{\prime\prime} \approx (\omega
t)^{-b^{\prime\prime}}$ where $b^{\prime\prime} = (d-\theta)/z(T)$, $d$ (=
3) is the dimension, and $\theta = 0.20 \pm 0.03$.  Experimentally, as
shown in our previous paper \cite{Suzuki2003},
$\Delta\chi^{\prime\prime}(\omega,t)$ at 3.75 K obeys the $\omega
t$-scaling law, $\Delta\chi^{\prime\prime}(\omega,t) \approx (\omega
t)^{-b^{\prime\prime}}$ with $b^{\prime\prime} = 0.255 \pm 0.005$, where
$\Delta\chi^{\prime\prime}(\omega,t) = \chi^{\prime\prime}(\omega,t)
-\langle\chi^{\prime\prime}_{0}(\omega)\rangle$, where
$\langle\chi^{\prime\prime}_{0}(\omega)\rangle$ is the stationary
frequency-dependent absorption.

\subsection{\label{backB}$S(t)$ and $\chi^{\prime\prime}(t_{w};t+t_{w})$}
Here we present a simple review on the relation between $S(t)$ and
$\chi^{\prime\prime}$.  The absorption $\chi^{\prime\prime}$ is evaluated
from the spin auto-correlation function $C(t_{a}-t;t_{a}) =
\overline{\langle S_{i}(t_{a}-t)S_{i}(t_{a})\rangle}$ using the
fluctuation-dissipation theorem (FDT) as \cite{Komori1999,Takayama2002}
\begin{equation} 
\chi^{\prime\prime}(\Delta t_{\omega};t+\Delta t_{\omega}) 
\approx (-\pi/2T) \partial C(\Delta t_{\omega};t+\Delta 
t_{\omega})/\partial\ln t, 
\label{eq:three} 
\end{equation} 
where $t_{a} = t+\Delta t_{\omega}$, $\Delta t_{\omega} = 2\pi/\omega$
(typically $\Delta t_{\omega} \leq 10^{2}$ sec) and $t$ is much larger than
$\Delta t_{\omega}$.  In the auto-correlation function, the over-line
denotes the average over sites and over different realizations of bond
disorder, and the bracket the average over thermal noises.  For slow
processes, the dispersion $\chi^{\prime}(\Delta t_{\omega};t+\Delta
t_{\omega})$ is approximated by
\begin{equation} 
\chi^{\prime}(\Delta t_{\omega};t+\Delta t_{\omega}) 
\approx [1-C(\Delta t_{\omega};t+\Delta t_{\omega})]/T.
\label{eq:four} 
\end{equation} 
In the quasi-equilibrium regime where the FDT holds, the ZFC 
susceptibility $\chi_{ZFC}$($t_{w}$\textit{; t}+$t_{w}$) is 
described by \cite{Komori1999,Picco2001}
\begin{equation} 
\chi_{ZFC}(t_{w};t+t_{w}) \approx [1-C(t_{w};t+t_{w})]/T, 
\label{eq:five} 
\end{equation} 
where $t_{a}=t+t_{w}$ and $t_{w}$ is a wait time: typically, $t_{w} \approx
10^{3} - 10^{5}$ sec.  Then the relaxation rate $S(t)$ is described by
\begin{eqnarray}
S(t) &=& {\rm d}\chi_{ZFC}(t_{w};t+t_{w}){\rm d}\ln t \nonumber\\
&=& (-1/T)\partial C(t_{w};t+t_{w})/\partial\ln t,
\label{eq:six} 
\end{eqnarray} 
which corresponds to $(2/\pi)\chi^{\prime\prime}(t_{w};t+t_{w})$
\cite{Lundgren1990}.

It is predicted that $C(t_{w};t+t_{w})$ can be decomposed into a stationary
part $C_{st}(t)$ and an aging part $C_{ag}(t_{w};t+t_{w})$
\cite{Yoshino2002}.  The latter is approximately described by a scaling
function of $L_{T}(t)$ and $R_{T}(t_{w})$.  The corresponding aging part of
$S(t)$ exhibits a peak at a characteristic time $t_{cr}$ ($\approx t_{w}$)
\cite{Yoshino2002,Andersson1992}, where $L_{T}(t) \approx R_{T}(t_{w})$,
showing a crossover between quasi-equilibrium region and non-equilibrium
region.

\section{\label{exp}Experimental procedure}
The DC magnetization and AC susceptibility of
Cu$_{0.5}$Co$_{0.5}$Cl$_{2}$-FeCl$_{3}$ GBIC were measured using a SQUID
magnetometer (Quantum Design, MPMS XL-5) with an ultra low field capability
option.  The remnant magnetic field was reduced to zero field (exactly less
than 3 mOe) at 298 K for both DC magnetization and AC susceptibility
measurements.  The AC magnetic field used in the present experiment has a
peak magnitude of $h = 0.1$ Oe and frequency $f = \omega/2\pi = 0.05$ Hz. 
Each experimental procedure for measurements is presented in the text and
figure captions.  The detail of sample characterizations and sample
preparation is given in the previous paper \cite{Suzuki2003}.

In our measurement of the time ($t$) dependence of the zero-field cooled
(ZFC) magnetization ($M_{ZFC}$), the time required for the ZFC aging
protocol and subsequent wait time was precisely controlled.  Typically it
takes $240 \pm 3$ sec to cool the system from 10 K to 3.75 K. It takes
another $t_{w0} = 230 \pm 3$ sec until $T$ (= 3.75 K) becomes stable within the
uncertainty ($\pm 0.01$ K).  The system is kept at $T = 3.75$ K and $H = 0$
for a wait time $t_{w}$ ($0 \leq t \leq 3 \times 10^{4}$ sec).  The time
for setting up a magnetic field from $H = 0$ to $H = 5$ Oe is $68 \pm 2$
sec.  In the ZFC measurement, the sample is slowly moved through the
pick-up coils over the scan length (4 cm).  The magnetic moment of the
sample induces a magnetic flux change in the pick-up coils.  It takes 12
sec for each scan.  The data at $t$ is regarded as the average of $M_{ZFC}$
measured over the scanning time $t_{s}$ between the times $t-(t_{s}/2)$ and
$t-(t_{s}/2)$.  Thus the time window $\Delta t$ is a scanning time
($t_{s}$).  The measurement was carried out at every interval of
$t_{s}+t_{p}$, where $t_{p}$ is a pause between consecutive measurements. 
Typically we used (i) the time window $\Delta t = 36$ sec for three scans
and $t_{p} = 45$ sec or 30 sec for $t_{w}$ ($\geq 2.0 \times 10^{3}$ sec),
and (ii) the time window $\Delta t = 12$ sec for one scan and $t_{p} = 1$
or 2 sec for either $t_{w}$ ($\leq 10^{3}$ sec) or $H \geq 100$ Oe.

\section{\label{result}Result}
\subsection{\label{resultA}$S(t):t_{w}$ and $T$ dependence}

\begin{figure}
\includegraphics[width=6.0cm]{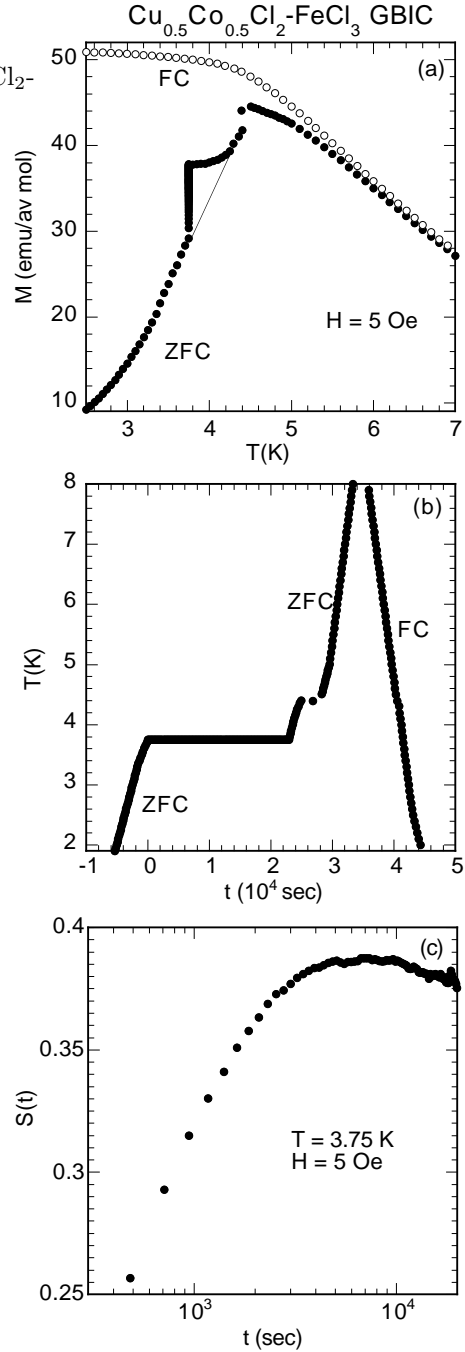}%
\caption{(a) $T$ dependence of $M_{ZFC}$ and $M_{FC}$ at $H = 5$ Oe for
Cu$_{0.5}$Co$_{0.5}$Cl$_{2}$-FeCl$_{3}$ GBIC. The measurement was carried
out after the ZFC aging protocol: annealing of the system at 50 K for $1.2
\times 10^{3}$ sec at $H = 0$ and quenching from 50 to 1.9 K. During the
ZFC measurement the system was aging at 3.75 K for $2.27 \times 10^{4}$
sec.  (b) The change of $T$ with $t$ during the measurement of $M_{ZFC}$
and $M_{FC}$.  The discontinuity of $T$ with $t$ at $T = 4.4$ K is due to
the system error occurring in the process of temperature stabilization. 
(c) The relaxation rate $S(t)$ ($= (1/H)$d$M_{ZFC}(t)/$d$\ln t$) at 3.75
K. The time taken during the measurement of $M_{ZFC}$ from 1.9 to 3.75 K at
$H$ = 5 Oe was $t_{0} = 5.4 \times 10^{3}$ sec.  $t = 0$ is a time just
after $T$ becomes 3.75 K.}
\label{fig:one}
\end{figure}

The $t$ dependence of $M_{ZFC}$ was measured under various conditions. 
Figure \ref{fig:one}(a) shows the $T$ dependence of the ZFC and FC
magnetization ($M_{ZFC}$ and $M_{FC}$) at $H$ = 5 Oe, where $H$ is applied
along a direction perpendicular to the c axis.  The system was quenched
from 50 to 1.9 K in the absence of $H$ before the measurement.  The change
of $T$ with the time $t$ during the measurement is shown in
Fig.~\ref{fig:one}(b).  The ZFC magnetization was measured with increasing
$T$ from 1.9 to 3.75 K in the presence of $H$ (= 5 Oe).  The system was
kept at $T = 3.75$ K for $2.27 \times 10^{4}$ sec.  Subsequently the ZFC
measurement was continued from 3.75 to 8 K. The system was annealed at 50 K
for $1.2 \times 10^{3}$ sec.  The FC magnetization was measured from 8 to
1.9 K. There is a remarkable increase of $M_{ZFC}$ during the one stop at
$T = 3.75$ K. When the warming up of the system was restarted, the increase
in $M_{ZFC}$ with $T$ is much weaker than that in $M_{ZFC}^{ref}$ as a
reference without the stop: $M_{ZFC}$ merges with $M_{ZFC}^{ref}$ well
above $T_{g}$.  Similar result of $M_{ZFC}$ vs $T$ was reported for a 3D
Ising SG Fe$_{0.5}$Mn$_{0.5}$TiO$_{3}$.  Bernardi et al. 
\cite{Bernardi2001} have concluded that $M_{ZFC}$ is described by a scaling
function of $R_{T}(t_{w})$ and $L_{T}(t)$ [$M_{ZFC}(T,t) =
G(L_{T}(t),R_{T}(t_{w}))$], where $G$ is the scaling function.  The $t$
dependence of $M_{ZFC}$ was monitored during the stop at $T = 3.75$ K,
where the time taken during the measurement from 1.9 to 3.75 K was $5.4
\times 10^{3}$ sec, corresponding to a wait time $t_{w}$.  Figure
\ref{fig:one}(c) shows the $t$ dependence of the relaxation rate $S(t)$,
where the origin of $t$ ($t = 0$) is a time when $T$ reaches 3.75 K.

\begin{figure}
\includegraphics[width=7.5cm]{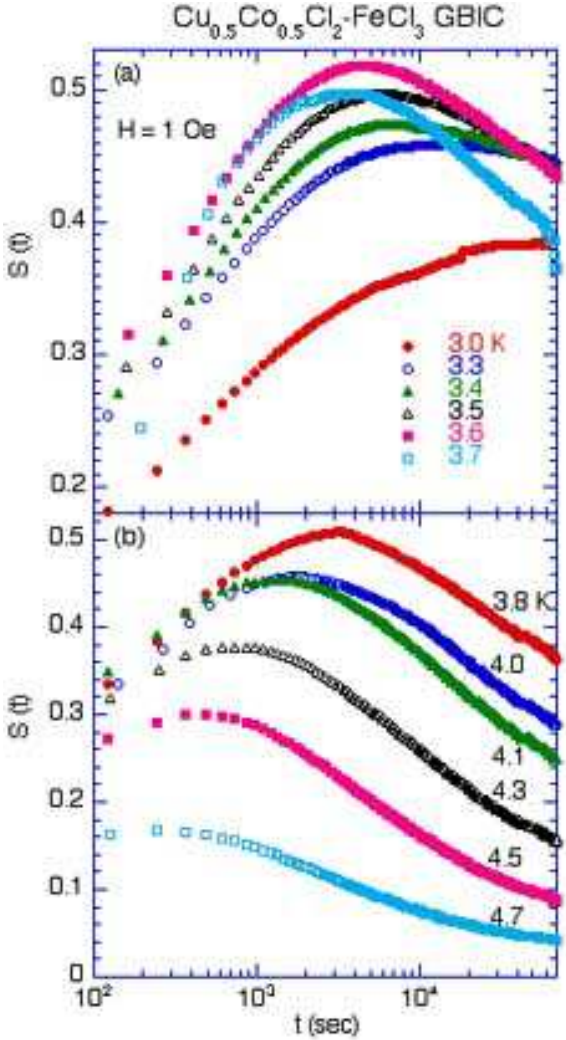}%
\caption{(a) and (b) $t$ dependence of $S$ for $3.0 \leq T \leq 4.7$ K. $H
= 1$ Oe.  The measurement of $\chi_{ZFC}$ vs $t$ was carried out after the
ZFC aging protocol (annealing of the system at 50 K and $H = 0$ for $1.2
\times 10^{3}$ sec, quenching from 50 K to $T$, and isothermal aging for
$t_{w} = 2.0 \times 10^{3}$ sec).  $t = 0$ is the time just after $H = 1$
Oe is applied at $T$.}
\label{fig:two}
\end{figure}

\begin{figure}
\includegraphics[width=7.5cm]{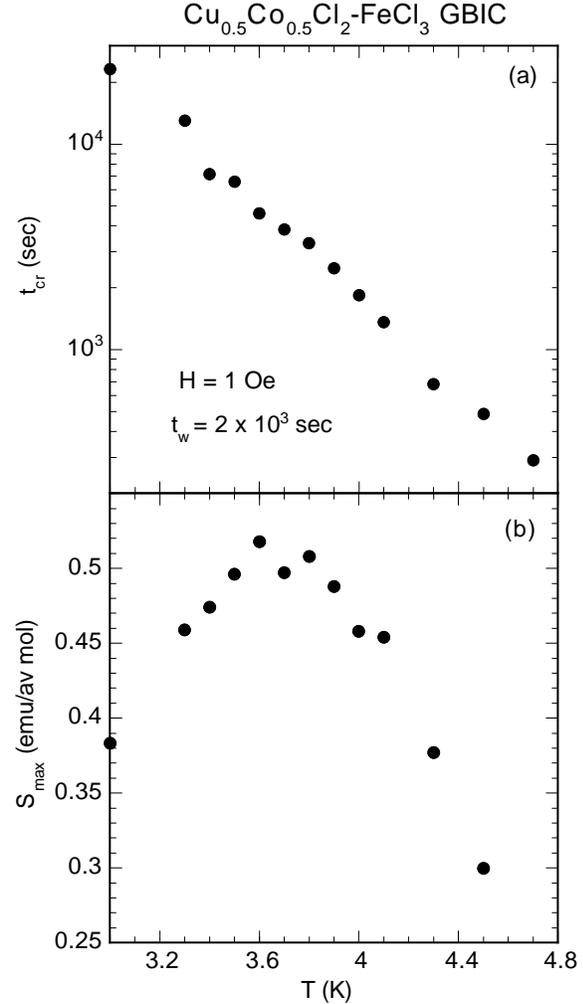}%
\caption{$T$ dependence of $t_{cr}$ at which $S(t)$ has a peak value
$S_{max}$.  $t_{w} = 2.0 \times 10^{3}$ sec.  $H = 1$ Oe.  (b) $T$
dependence of the peak height $S_{max}$ (see Fig.~\ref{fig:two}).}
\label{fig:three}
\end{figure}

We have measured the $t$ dependence of $\chi_{ZFC}$ ($= M_{ZFC}/H$) at
various fixed $T$, where $t_{w} = 2.0 \times 10^{3}$ sec and $H = 1$ Oe. 
The measurement was carried out after the ZFC aging protocol: annealing of
the system at $T = 50$ K and $H = 0$ for $1.2 \times 10^{3}$ sec, quenching
from 50 K to $T$ at $H = 0$, and isothermal aging at $T$ for $t_{w}$.  The
origin of $t$ ($t$ = 0) is a time just after $H$ = 1 Oe is applied at $T$. 
We find that $M_{ZFC}$ increases with increasing $t$, depending on $T$. 
Figures \ref{fig:two}(a) and (b) show the $t$ dependence of $S$ at various
$T$ ($3.0 \leq T \leq 4.7$ K), where $H = 1$ Oe and $t_{w} = 2.0 \times
10^{3}$ sec.  Each $S(t)$ curve exhibits a peak at a characteristic time
$t_{cr}$, shifting to the short $t$-side with increasing $T$.  The peak
width becomes much broader with decreasing $T$.  Figures \ref{fig:three}(a)
and (b) show the $T$ dependence of $t_{cr}$ and the peak height $S_{max}$,
respectively, which are derived from Figs.~\ref{fig:two}(a) and (b).  The
characteristic time $t_{cr}$ increases with decreasing $T$.  The $T$
dependence of $\ln(t_{cr}/t_{w})$ will be discussed in Sec.~\ref{dis} in
association with the scaling relations.  The peak height $S_{max}$ exhibits
a broad peak between 3.6 and 3.8 K just below $T_{g}$.  Similar behavior of
$S$ vs $t$ at various $T$ ($<T_{g}$) has been reported by Zotev et al.  for
Cu (1.5 at \% Mn) \cite{Zotev2003}, where $S(t)$ is defined from the
thermoremnant magnetization $M_{TR}$ instead of $M_{ZFC}$.  The relaxation
rate $S(t)$ exhibits a peak at $t = t_{cr}$, whose peak width becomes much
broader with decreasing $T$.  The peak height $S_{max}$ decreases with
decreasing $T$ below $T_{g}$.

\begin{figure}
\includegraphics[width=6.5cm]{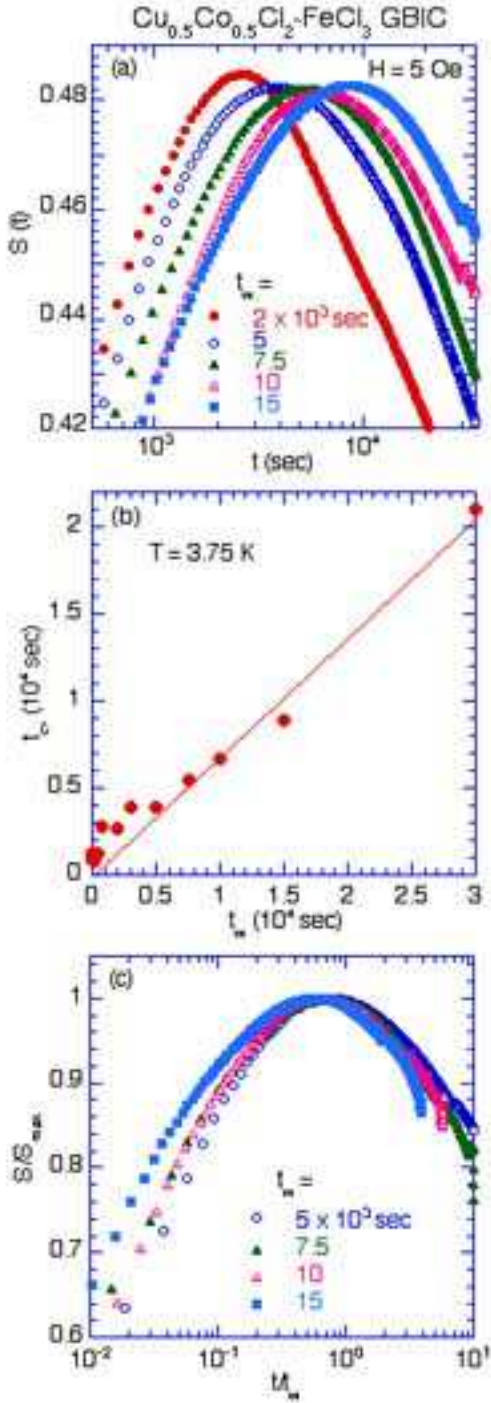}%
\caption{(a) $t$ dependence of $S$ for $0 \leq t_{w} \leq 1.5 \times
10^{4}$ sec.  $T = 3.75$ K. $H = 5$ Oe.  The ZFC aging protocol: annealing
of the system at 50 K for $1.2 \times 10^{3}$ sec at $H = 0$, quenching
from 50 to 3.75 K, and then isothermal aging at 3.75 K and $H = 0$ for a
wait time $t_{w}$.  The measurement was started at $t = 0$ when the field
$H$ is turned on.  (b) $t_{cr}$ vs $t_{w}$.  The straight line denotes a
relation given by $t_{cr} = (0.68 \pm 0.03)t_{w}$.  (c) Scaling plot of
$S/S_{max}$ vs $t/t_{w}$ for the limited $t_{w}$ and $t$ ($5.0 \times 10^{3} \leq
t_{w} \leq 1.5 \times 10^{4}$ sec, $0 \leq t \leq 6.0 \times 10^{4}$
sec).}
\label{fig:four}
\end{figure}

Figure \ref{fig:four}(a) shows the $t$ dependence of $S$ at various 
$t_{w}$, where $T = 3.75$ K
and $H = 5$ Oe.  The relaxation rate $S(t)$ shows a peak at $t = t_{cr}$,
shifting to the long-$t$ side with increasing $t_{w}$.  Similar behavior of
$S(t)$ vs $t$ at various $t_{w}$ has been observed by J\"{o}nsson et al. 
for Ag (11 at.\% Mn) \cite{Jonsson2002b}, where $t_{cr} \approx t_{w}$.  In
Fig.~\ref{fig:four}(b) we show the characteristic time $t_{cr}$ as a
function of $t_{w}$ for $0 \leq t_{w} \leq 3.0 \times 10^{4}$ sec, where a
straight line denotes the relation described by $t_{cr} = (0.68 \pm
0.03)t_{w}$.  Figure \ref{fig:four}(c) shows the plot of $S(t)/S_{max}$ as
a function of $t/t_{w}$, where only the data of $S$ vs $t$ with long
$t_{w}$ ($5 \times 10^{3} \leq t_{w} \leq 1.5 \times 10^{4}$ sec) for $0 <
t < 6 \times 10^{4}$ sec are used.  It seems that $S(t)/S_{max}$ is well
described by a scaling function of $t/t_{w}$ [$S(t)/S_{max} = F(t/t_{w})$]
in the region of long $t_{w}$, although the data at $t_{w} = 1.5 \times
10^{4}$ sec slightly deviates from the other data.  The scaling function
$F(x)$ has a very broad peak centered at $x \approx 0.68$.  This result
suggests that the spin auto-correlation function $C(t_{w},t+t_{w})$ is
described by a scaling function of only $t/t_{w}$, since it is related to
$S(t)$ by Eq.(\ref{eq:six}).

\subsection{\label{resultB}$S(t)$ for $0 \leq t_{w} \leq 750$ sec}

\begin{figure}
\includegraphics[width=7.5cm]{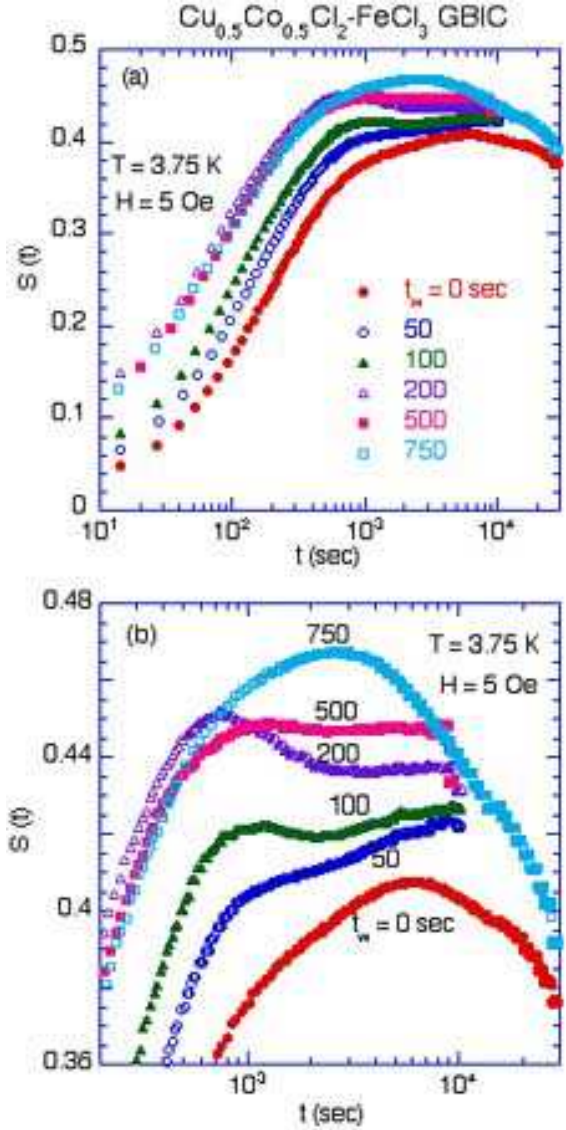}%
\caption{(a) and (b) $t$ dependence of $S$ for $0 \leq t_{w} \leq 750$ sec. 
$T = 3.75$ K. $H = 5$ Oe.}
\label{fig:five}
\end{figure}

\begin{figure}
\includegraphics[width=7.5cm]{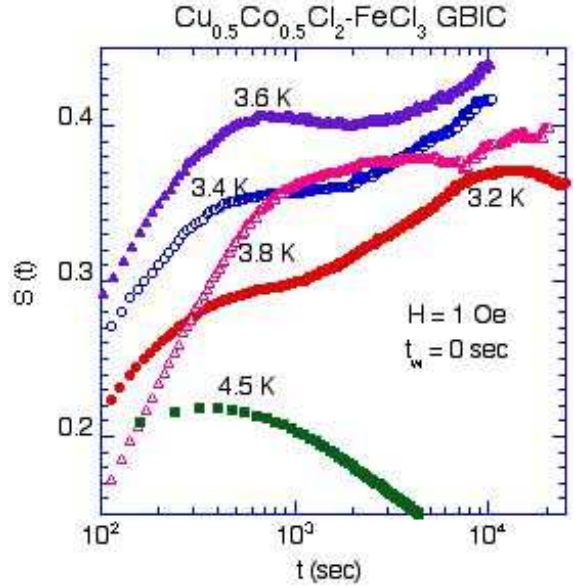}%
\caption{$t$ dependence of $S$ at various $T$.  $t_{w} = 0$ sec.  $H = 1$
Oe.}
\label{fig:six}
\end{figure}

We have measured the $t$ dependence of $\chi_{ZFC}$ at $T = 3.75$ K and $H
= 5$ Oe as a wait time $t_{w}$ ($0 \leq t_{w} \leq 750$ sec) is varied as a
parameter.  The measurement was carried out after the ZFC aging protocol:
annealing of the system at 50 K for $1.2 \times 10^{3}$ sec at $H = 0$,
quenching from 50 to 3.75 K, and then isothermal aging at $T = 3.75$ K and
$H = 0$ for a wait time $t_{w}$.  The origin of $t$ ($t = 0$) is a time
just after the field $H$ is turned on.  As is described in Sec.~\ref{exp},
it takes $t_{w0} = 230 \pm 3$ sec until the temperature becomes stable at
3.75 K within the experimental uncertainty of $\pm 0.01$ K, after quenching
the system from 50 to 3.75 K at $H = 0$.  This time $t_{w0}$ is not
included in the wait time $t_{w}$.  If $t_{w0}$ is included in $t_{w}$,
however, the effective wait time may be longer than $t_{w}^{eff}$ ($=
t_{w0}+t_{w}$).  If this is the case, the measurement with $t_{w} = 0$ is
not possible in a strict sense.  In spite of such a situation, here we
assume that $t_{0}$ is not included in $t_{w}$.  Figures \ref{fig:five}(a)
and (b) show the $t$ dependence of $S$ for $0 \leq t_{w} \leq 750$ sec,
where $T = 3.75$ K and $H = 5$ Oe.  For $t_{w} = 0$, $S(t)$ shows a
shoulder around $t = 10^{3}$ sec and a broad peak at $t_{cr} = 6.45 \times
10^{3}$ sec.  For $t_{w} = 100$ sec, a small peak of $S(t)$ is observed at
$t_{cr} = 1.06 \times 10^{3}$ sec in addition to a possible broad peak
around $t_{cr} = 1.0 \times 10^{4}$ sec.  This small peak shifts to the
short $t$-side with increasing $t_{w}$: $t_{cr} = 740$ sec for $t_{w} =
200$ sec.  For $t_{w} = 750$ sec, a broad peak is observed at $t_{cr} = 2.8
\times 10^{3}$ sec.  The values of $t_{cr}$ thus obtained are also plotted
as a function of $t_{w}$ in Fig.~\ref{fig:four}(b).  It should be noted
that that for very short $t_{w}$ ($0 \leq t \leq 200$ sec), there are at
least two kinds of domains: domains with large size corresponding to long
$t_{cr}$ ($\approx 6.5 \times 10^{3}$ sec) coexist with domains with small
size corresponding to short $t_{cr}$ ($\approx 750$ sec) in the regular
aging regime.  Figure \ref{fig:six} shows the $t$ dependence of $S(t)$ at
various $T$ ($3.4 \leq T \leq 4.5$ K) and $H = 1$ Oe for $t_{w} = 0$.  At
$T = 3.2$ K, $S(t)$ shows a shoulder around $t = 400$ sec and a very broad
peak at $t_{cr} = 7.1 \times 10^{3}$ sec.  This shoulder shifts to the long
$t$-side with increasing $T$ and changes in to a broad peak.  At $T = 3.6$
K, $S(t)$ shows a peak at $t_{cr} = 710$ sec.  It tends to increase with
further increasing $t$, suggesting that $t_{cr}$ is longer than $10^{4}$
sec.  At $T = 3.8$ K, $S(t)$ shows a very broad peak centered at $t_{cr} =
3.8 \times 10^{3}$ sec.

Similar behavior has been reported by Rodriguez et al. 
\cite{Rodriguez2003} in the time decay of the thermal remnant
magnetization (TRM) of Cu$_{0.94}$Mn$_{0.06}$ ($T_{g} = 31.5$ K) with
various wait time ($t_{w}$ = 0 - $10^{4}$ sec) and a series of rapid FC cooling
protocol from 35 to 26 K at $H$ (= 20 Oe).  The relaxation rate $S(t)
[=-(1/H)$d$M_{TRM}(t)/$d$\ln t]$ for $M_{TRM}(t)$ is equivalent to $S(t)$
[$=-(1/H)$d$M_{ZFC}(t)/$d$\ln t$] for $M_{ZFC}(t)$.  They have shown that
$S(t)$ at $t_{w} = 0$ exhibits a broad peak at an effective time
$t_{c}^{eff}$ (= 19 - 406 sec), which is strongly dependent on the FC
cooling protocol.  For the larger $t_{c}^{eff}$, there is a significant
contamination in $S(t)$ at $t_{w}$ ($\neq 0$) from the FC cooling protocol. 
Recently the long-time decay of $M_{TRM}(t)$ has been also examined by
Kenning et al.  \cite{Kenning2004} for the same system with rapid FC
cooling protocol and short wait time ($t_{w}$ = 7 - 110 sec).  A
$t_{w}$-independent long-time decay overlaps with the $t_{w}$-dependent
short-time decay.  For the short-time decay, the corresponding $S(t)$
exhibits a peak at $t_{cr}¥$ ($\approx t_{w}$).  The long-time decay may be
related to the initial state distribution developed during the FC
cooling protocol.

\subsection{\label{resultC}$S(t)$ under the $H$-shift}

\begin{figure}
\includegraphics[width=7.5cm]{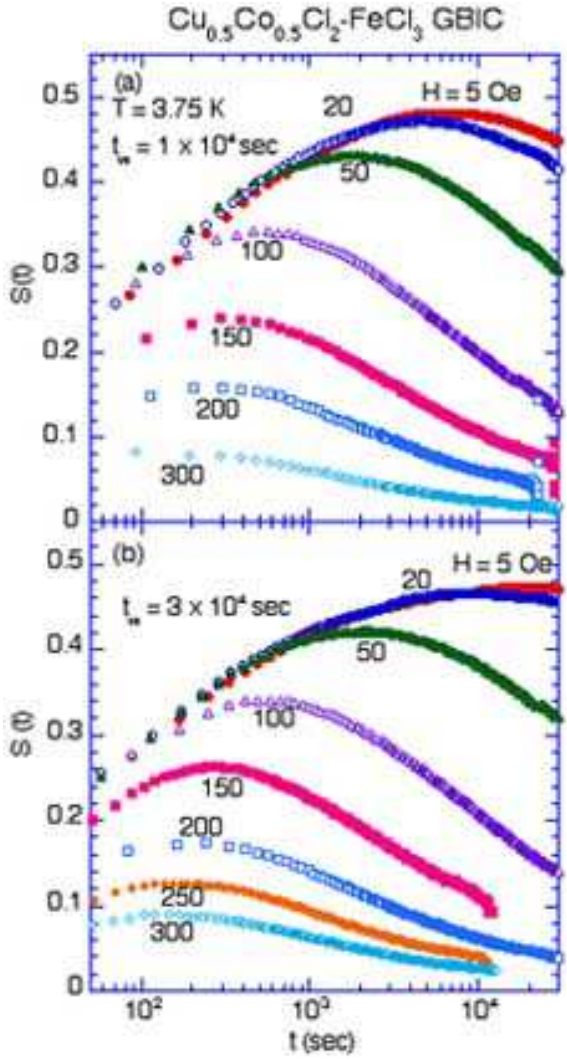}%
\caption{$t$ dependence of $S$ for $5 \leq H \leq 300$ Oe.  $T = 3.75$ K.
(a) $t_{w} = 1.0 \times 10^{4}$ sec.  (b) $t_{w} = 3.0 \times 10^{4}$ sec. 
The time $t = 0$ is a time when $H$ is turned on.  The ZFC aging protocol
before the measurement is similar to that used in Fig.~\ref{fig:two}.}
\label{fig:seven}
\end{figure}

\begin{figure}
\includegraphics[width=7.5cm]{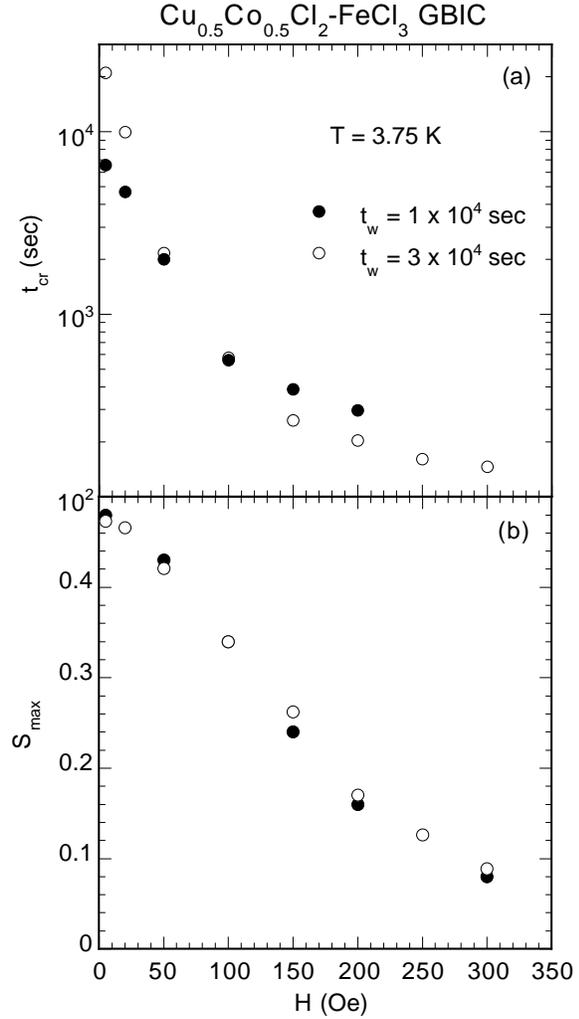}%
\caption{$H$ dependence of (a) $t_{cr}$ and (b) $S_{max}$ for $t_{w} = 1.0
\times 10^{4}$ and $3.0 \times 10^{4}$ sec, which is obtained from
Fig.~\ref{fig:seven}.}
\label{fig:eight}
\end{figure}

We have measured the $t$ dependence of $\chi_{ZFC}$ at $T = 3.75$ K for
various $H$, where $t_{w}=1.0 \times 10^{4}$ and $3.0 \times 10^{4}$ sec. 
The measurements were carried out after the ZFC aging protocol: annealing
of the system at $T = 50$ K and $H = 0$ for $1.2 \times 10^{3}$ sec,
quenching from 50 to 3.75 K at $H = 0$, and isothermal aging at 3.75 K for
$t_{w}$.  The origin of $t$ ($t = 0$) is the time just after $H$ is turned
on.  The window time used in this measurement was 12 sec.  Figures
\ref{fig:seven}(a) and (b) show the $t$ dependence of $S$ for $t_{w} = 1.0
\times 10^{4}$ sec and $3.0 \times 10^{4}$ sec as $H$ is varied as a
parameter, where $T = 3.75$ K. The relaxation rate $S(t)$ exhibits a peak
at $t = t_{cr}$, corresponding to a characteristic time scale of crossover
from the isothermal aging state under $H = 0$ to that under a finite $H$. 
Figure \ref{fig:eight}(a) shows the plot of $t_{cr}$ at $T = 3.75$ K as a
function of $H$ for $t_{w} =1.0 \times 10^{4}$ and $3.0 \times 10^{4}$ sec. 
The value of $t_{cr}$ for $t_{w} = 1.0 \times 10^{4}$ and $3.0 \times
10^{4}$ sec drastically decreases with increasing $H$.  The value of
$t_{cr}$ for $t_{w} = 3 \times 10^{4}$ sec is much larger than that for
$t_{w} = 1 \times 10^{4}$ sec at low $H$ ($H < 50$ Oe).  However, they are
almost identical irrespective of $t_{w}$ for high $H$ ($H > 100$ Oe).  We
note that $\ln t_{cr}$ is linearly dependent on $H$ only for $H <
H_{\alpha}$: $H_{\alpha} \approx 50$ Oe for $t_{w} = 3.0 \times 10^{4}$ sec
and 100 Oe for $t_{w} = 1.0 \times 10^{4}$ sec.  In Fig.~\ref{fig:eight}(b)
we show the $H$ dependence of $S_{max}$.  The peak value $S_{max}$ linearly
decreases with increasing $H$ irrespective of $t_{w}$ and tends to reduce
to zero above 300 Oe.  Similar behavior of $t_{cr}$ vs $H$ has been
reported by Zotev et al.  \cite{Zotev2002} in Cu (1.5 at.\% Mn).  The break
of the linear dependence of $\ln t_{cr}$ on $H$ occurs at $H_{\alpha}$. 
The value of $H_{\alpha}$ decreases with increasing $t_{w}$.  Above
$H_{\alpha}$ the data of $\delta M$ ($= M_{TRM}-M_{FC}+M_{ZFC}$) vs $H$
greatly deviates from zero, where $M_{TRM}$ is the thermal remnant
magnetization.  This result suggests that the crossover from the quasi
equilibrium to the nonequibrium regime occurs at $H_{\alpha}$, leading to
the violation of FDT.

According to Takayama \cite{Takayama2003}, the $H$ dependence of $t_{cr}$
under the $H$-shift aging process is governed by three lengths,
$L_{T}(t_{cr},H)$, $R_{T}(t_{w})$, and the crossover length $L_{H}$
\cite{Fisher1988} given by
\begin{equation} 
L_{H}/L_{0} \approx (H/\Upsilon_{H})^{-1/\delta}, 
\label{eq:seven} 
\end{equation} 
where $\delta = (d/2-\theta)$ and $\Upsilon_{H}$ is the magnetic field
corresponding to a wall stiffness $\Upsilon$ (a typical energy setting the
scale of free energy barriers between conformations).  The scaling relation
is predicted to exist between the normalized lengths $y =
L_{T}(t_{cr},H=0)/L_{H}$ and $x = R_{T}(t_{w})/L_{H}$: $y = x-c_{H}
x^{1+\delta}$ with $c_{H} = 0.15$ for $x<1$.  For $x \ll 1$ corresponding
to the case of low $H$ and short $t_{w}$, $y = x$, implying that $t_{cr} =
t_{w}$.  For $x > 0.2$ corresponding to large $H$ and long $t_{w}$, the
curve $y(x)$ deviates below the line $y = x$, indicating that $t_{cr}$ is
shorter than $t_{w}$.  From this scaling relation, $t_{cr}$ can be obtained
as
\begin{equation} 
t_{cr}=t_{w}[1-c_{H}(H/\Upsilon_{H})(t_{w}/t_{0})^{\delta/z(T)}]^{z(T)}. 
\label{eq:eight} 
\end{equation} 
The following features of $t_{cr}$ vs $H$ are derived from
Eq.(\ref{eq:eight}).  (i) The time $t_{cr}$ is not simply proportional to
$t_{w}$.  (ii) In the limit of $H \approx 0$, $\ln t_{cr}$ is linearly
dependent on $H$: 
\[
\ln(t_{cr}/t_{w}) = -\alpha_{H}H,
\]
with 
\[
\alpha_{H} =
z(T)(c_{H}/\Upsilon_{H}) (t_{w}/t_{0})^{\delta/z(T)}.  
\]
The slope
$\alpha_{H}$ increases with increasing $t_{w}$.  (iii) The $T$ dependence
of $t_{cr}$ comes from the exponent $z(T)$.  The slope $\alpha_{H}$
increases with increasing $T$ below $T_{g}$ mainly because of the factor
($t_{w}/t_{0})^{\delta/z(T)}$.

We find that our results of Fig.~\ref{fig:eight}(a) is consistent with the
above predictions.  The slope ($\alpha_{H} = 0.0505 \pm 0.0003$ /Oe) for
$t_{w} = 3.0 \times 10^{4}$ sec is larger than that ($\alpha_{H} = 0.0262
\pm 0.0005$ /Oe) for $t_{w} = 1.0 \times 10^{4}$ sec.  If we assume that
$\delta = 1.37$, $1/z(T) = bT/T_{g}$, $T_{g} = 3.92$ K, $b = 0.16$, and
$t_{0} = \tau^{*} = 5.29 \times 10^{-6}$ sec \cite{Suzuki2003}, then the
value of $\Upsilon_{H}$ can be estimated as $\Upsilon_{H} = 2.15$ kOe for
$t_{w} = 3.0 \times 10^{4}$ sec and $\Upsilon_{H} = 3.30$ kOe for $t_{w} =
1.0 \times 10^{4}$ sec.  Our numerical calculation of $\alpha_{H}$ for 3 K
$\leq T \leq T_{g}$ predicts that the slope $\alpha_{H}$ increases with
increasing $T$ below $T_{g}$: $\alpha_{H} = 0.036$ at $T = 3.4$ K and
$\alpha_{H} = 0.044$ at $T = 3.6$ K for $t_{w} = 3.0 \times 10^{4}$ sec. 
Although the measurement on $\alpha_{H}$ vs $T$ has not been carried out in
the present system, this prediction is consistent with the result 
reported by Zotev et al. 
\cite{Zotev2002} in Cu (1.5 at.\% Mn) as the increase of $\alpha_{H}$ vs
$T$ with increasing $T$ for $0.7 < T/T_{g} < 0.85$.

\subsection{\label{resultD}$S(t)$ under the $T$-shift}

\begin{figure}
\includegraphics[width=6.5cm]{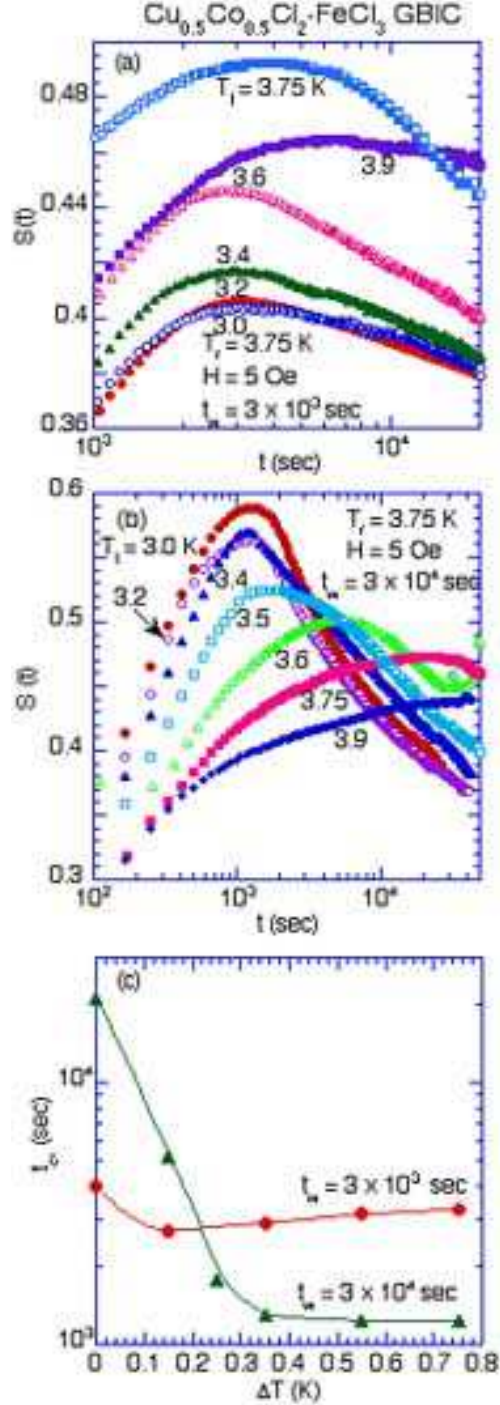}%
\caption{$t$ dependence of $S$ at $T_{f} = 3.75$ K under the $T$ shift
from $T_{i}$ to $T_{f}$, where $T_{i} = 3.0 - 3.9$ K. (a) $t_{w} = 3.0
\times 10^{3}$ sec.  (b) $t_{w} = 3.0 \times 10^{4}$ sec.  The ZFC aging
protocol is as follows: quenching of the system from 50 K to $T_{i}$, and
isothermal aging at $T = T_{i}$ and $H = 0$ for $t_{w}$.  The measurement
was started at $t = 0$, just after $T$ was shifted from $T_{i}$ to $T_{f} =
3.75$ K and subsequently $H$ (= 5 Oe) was turned on.  (c) $t_{cr}$ vs
$\Delta T$ for the case of positive $T$-shift ($t_{w} = 3.0 \times 10^{3}$
and $3.0 \times 10^{4}$ sec), where $T_{i}=T_{f}-\Delta T$ ($\Delta T > 0$)
and $T_{f} = 3.75$ K. The solid lines are guides to the eye.}
\label{fig:nine}
\end{figure}

We have measured the $t$ dependence of $\chi_{ZFC}$ under the $T$-shift
from the initial temperature $T_{i}$ to the final temperature $T_{f}$ (=
3.75 K $= 0.957 T_{g}$), where $T_{i}$ = 3, 3.2 3.4, 3.5, 3.6, and 3.9 K.
The measurement was carried out after the ZFC aging protocol: quenching of
the system from 50 K to $T_{i}$, and isothermal aging at $T = T_{i}$ and $H
= 0$ for $t_{w}$ ($= 3.0 \times 10^{3}$ and $3 \times 10^{4}$ sec).  The
origin of time ($t = 0$) is a time just after $T$ was shifted from $T_{i}$
to $T_{f} = 3.75$ K and subsequently $H$ (= 5 Oe) was turned on.  Figures
\ref{fig:nine}(a) and (b) show the $t$ dependence of $S$ at $T_{f} = 3.75$
K at various initial temperature $T_{i}$.  The temperature difference is
defined as $\Delta T = T_{f}-T_{i}$: the positive $T$-shift for $\Delta T >
0$ and negative $T$-shift for $\Delta T < 0$.  The relaxation rate $S(t)$
exhibits a peak at $t = t_{cr}$ irrespective of the sign of $\Delta T$. 
The width of the peak in $S(t)$ for the negative $T$-shift is much broader
than that for the positive $T$-shift.  In Fig.~\ref{fig:nine}(c) we show
$t_{cr}$ as a function of $\Delta T$ for the positive $T$-shift for $t_{w}
= 3.0 \times 10^{4}$ and $3.0 \times 10^{3}$ sec.  The decrease of $t_{cr}$
with increasing $\Delta T$ is observed at low $\Delta T$ for both $t_{w} =
3.0 \times 10^{4}$ and $3.0 \times 10^{3}$ sec.  The value of $t_{cr}$
becomes independent of $\Delta T$ at large $\Delta T$.  For the negative
$T$-shift, on the other hand, the value of $t_{cr}$ for $\Delta T = -0.15$
K ($T_{i} = 3.9$ K) is larger than that for $\Delta T = 0$ K ($T_{i} =
3.75$ K) for both $t_{w} = 3.0 \times 10^{4}$ and $3.0 \times 10^{3}$
sec.  Similar behaviors of $t_{cr}$ vs $\Delta T$ for the positive
$T$-shift have been reported by Granberg et al.  [Cu (10 at.~\%Mn)]
\cite{Granberg1988}, Djurberg et al.  [Cu (13.5 at.~\% Mn)]
\cite{Djurberg1999}, and Jonsson et al.  [Ag (11 at.~\% Mn)]
\cite{Jonsson1999}.

The temperature chaos scenario postulates that the SG equilibrium 
configurations at different temperatures at $T_{f}$ and $T_{i}$ 
are strongly correlated only up to the overlap-length $L_{\Delta T}$, 
beyond which these correlations decay to zero. From the droplet 
theory, the overlap length $L_{\Delta T}$ is described by 
\cite{Fisher1988,Komori1999}
\begin{equation} 
L_{\Delta T}/L_{0} \approx (T^{1/2}\vert\Delta T\vert/\Upsilon_{T}^{3/2})^{-1/\zeta}, 
\label{eq:nine} 
\end{equation} 
where $\zeta$ is the chaos exponent ($\zeta = d_{s}/2 -\theta$), $d_{s}$ is
the fractal dimension of the surface of the droplet, and $\Upsilon_{T}$ is
the temperature corresponding to the wall stiffness $\Upsilon$.  From this
scenario a scaling relation is predicted to exist between the normalized
lengths $y =[L_{Tf}(t_{cr})/L_{\Delta T}]$ and $x =
[R_{T_{i}}(t_{w})/L_{\Delta T}]$: $y = x-c_{T}x^{1+\zeta}$ with $c_{T} = 0.25$
and $\zeta = 0.385$ for $x < 0.15$ \cite{Jonsson2002b,Jonsson2003}.  The
cumulative aging corresponds to the relation $y = x$ which is valid in the
limit $x \approx 0$ (the small $\vert\Delta T\vert$ and short $t_{w}$). 
The large $\vert\Delta T\vert$ and long $t_{w}$ corresponds to large $x$. 
For $x > 0.05$, the curve $y$ deviates from the straight line $y = x$,
corresponding to a rejuvenation due to the temperature chaos effect
\cite{Jonsson2002b,Jonsson2003}.  The value of $t_{cr}/t_{0}$ can be
described by
\begin{equation} 
t_{cr}/t_{0}=(t_{w}/t_{0})^{T_{i}/T_{f}}[1-c_{T}p(T_{f})\vert\Delta 
T\vert(t_{w}/t_{0})^{\zeta/z(T_{i})}]^{z(T_{f})}, 
\label{eq:ten} 
\end{equation} 
with $p(T_{f}) = T_{f}^{1/2}/\Upsilon_{T}^{3/2}$.  In the limit of
$\vert\Delta T\vert \approx 0$, $\ln t_{cr}$ is linear to $T$:
\[
\ln(t_{cr}/t_{w}) = -\alpha_{T}\vert\Delta T\vert,
\]
with 
\[
\alpha_{T} =
z(T_{f}) c_{T} p(T_{f}) (t_{w}/t_{0})^{\zeta/z(T_{i})}.  
\]
In fact, the
prediction that $\alpha_{T}$ increases with increasing $t_{w}$ is
experimentally confirmed from Fig.~\ref{fig:nine}(c).  The curve ($\ln
t_{cr}$ vs $\Delta T$) is linearly dependent on $\Delta T$ ($> 0$) for
$\Delta T < \Delta T_{\alpha}$ with the slope $\alpha_{T}$: $\alpha_{T} =
9.9 \pm 0.4$ and $\Delta T_{\alpha} =0.3$ K for $t_{w} = 3.0 \times 10^{4}$
sec and $\alpha_{T} = 2.6$ and $\Delta T_{\alpha} \approx 0.15$ K for
$t_{w} = 3.0 \times 10^{3}$ sec.  If we assume that $\zeta = 0.385$, $c_{T}
= 0.25$, $1/z(T_{f}) = bT_{f}/T_{g}$, $b = 0.16$, and
$t_{0} = \tau^{*} = 5.29 \times 10^{-6}$ sec \cite{Suzuki2003} for $t_{w} =
3.0 \times 10^{4}$ sec, we have $p(T_{f}) = 1.62 \pm 0.07$ or $\Upsilon_{T}
= 1.12 \pm 0.03$ K for $t_{w} = 3.0 \times 10^{4}$ sec.

\begin{figure}
\includegraphics[width=7.5cm]{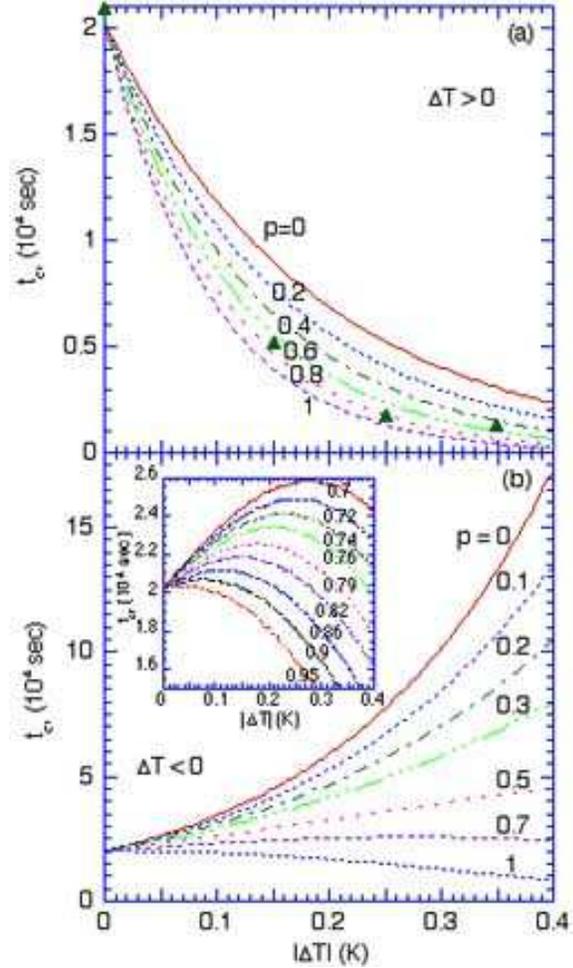}%
\caption{Numerical calculations of $t_{cr}$ vs $\vert\Delta T\vert$ given
by Eq.(\ref{eq:ten}) for $T_{f} = 3.75$ K, $T_{f} = T_{i}+\Delta T$, and
$t_{w} = t_{w}$(cal) $= 2.04 \times 10^{4}$ sec, where $T_{g} = 3.92$ K, $b
= 0.16$, $\zeta = 0.385$, $c_{T} = 0.25$, $t_{0} = \tau^{*} = 5.29 \times
10^{-6}$ sec, and $0 \leq p(T_{f}) \leq 1$.  (a) The positive $T$-shift
($\Delta T > 0$).  For comparison, our data of $t_{cr}$ vs $\Delta T$ 
for $t_{w} = 3.0 \times 10^{4}$ sec are denoted by solid triangles. 
The value of $t_{w}$ ($= t_{w}$(cal) $= 2.04 \times 10^{4}$ sec) used in
the calculation is different from that used in the experiment [$t_{w}$(exp)
$= 3.0 \times 10^{4}$ sec], where $t_{w}$(cal) $= 0.68t_{w}$(exp) (see
Fig.~\ref{fig:four}(b)).  (b) The negative $T$-shift ($\Delta T < 0$). 
Although the maximum of $\vert\Delta T\vert$ is $T_{g}-T_{f} = 0.17$ K in
the present system, the value of $t_{cr}$ for the negative $T$-shift is
independent of $T_{g}$ and is applicable to the other systems with 
different $T_{g}$.}
\label{fig:ten}
\end{figure}

Numerical calculations of Eq.(\ref{eq:ten}) are carried out as a function
of $\vert\Delta T\vert$ for $t_{w} = t_{w}$(cal) $= 2.04 \times 10^{4}$
sec, $T_{f} = 3.75$ K, $T_{i}= T-\Delta T$, $\zeta = 0.385$, $t_{0} =
\tau^{*} = 5.29 \times 10^{-6}$ sec, and $c_{T} = 0.25$.  Note that we use
$t_{w}$(cal) as $t_{w}$ instead of $t_{w}$(exp) ($= 3.0 \times 10^{4}$ sec)
[$t_{w}$(cal) $= 0.68t_{w}$(exp)], which leads to the better agreement
between our data and calculations at low $\vert\Delta T\vert$.  In
Figs.~\ref{fig:ten}(a) and (b) we show the results of $t_{cr}$ vs
$\vert\Delta T\vert$ for the positive and negative $T$-shifts,
respectively, where $p(T_{f})$ is varied as a parameter from 0 to 2.0.  For
comparison, our data of $t_{cr}$ vs $\Delta T$ with $t_{w} = 3.0 \times
10^{4}$ sec for the positive $T$-shift are also plotted in
Fig.~\ref{fig:ten}(a).  For the positive $T$-shift, $t_{cr}$ decreases with
increasing $\Delta T$, independent of $p(T_{f})$ for $0 \leq p(T_{f}) \leq
2$.  For the negative $T$-shift, $t_{cr}$ increases with increasing
$\vert\Delta T\vert$ for $0 \leq p(T_{f}) < 0.7$.  It increases with
increasing $\vert\Delta T\vert$, showing a peak, and decreases with further
increasing $\vert\Delta T\vert$ for $0.7 < p(T_{f}) < 0.92$ (see the inset
of Fig.~\ref{fig:ten}(b)).  On the other hand, it decreases with increasing
$\vert\Delta T\vert$ for $0.92 \leq p(T_{f}) < 2$.  This indicates that
$t_{cr}$ decreases with increasing $\vert\Delta T\vert$ for both the
positive and negative $T$-shifts for $0.92 \leq p(T_{f}) < 2$ (the
symmetric $T$-chaos).  We find that our data for both the positive and
negative $T$-shifts agree well with the curve with $p(T_{f}) \approx 0.8$
in Fig.~\ref{fig:ten}(a), corresponding to $\Upsilon_{T} = 1.80$ K. This
value of $\Upsilon_{T}$ is on the same order as that derived from the slope
$\alpha_{T}$.

For Ag (11 at.  \% Mn) \cite{Jonsson2002b}, $t_{cr}$ decreases with
increasing $\Delta T$ for the positive $T$-shift.  For the negative
$T$-shift, however, $t_{cr}$ shifts to the long-$t$ side at $\vert\Delta
T\vert = 0.1$ K and then $t_{cr}$ decreases with further increasing
$\vert\Delta T\vert$.  Similar behavior has been also observed in Cu (13.5
at.  \% Mn) \cite{Djurberg1999}, where $T_{f} = 58$ K and $t_{w} = 3.0
\times 10^{3}$ sec.  For the positive $T$-shift, $t_{cr}$ decreases with
increasing $\Delta T$ ($0 \leq \Delta T \leq 1.21$ K).  For the negative
$T$-shift, $t_{cr}$ increases with increasing $\vert\Delta T\vert$ for $0
\leq \vert\Delta T\vert \leq 0.6$ K, $t_{cr}$ becomes the longest at
$\vert\Delta T\vert = 0.6$ K, and $t_{cr}$ decreases with further
increasing $\vert\Delta T\vert$ for $0.6 \leq \vert\Delta T\vert \leq 4.03$
K. These behaviors can be well explained in terms of the scaling relations.

\subsection{\label{resultE}Memory effect in $\chi^{\prime}$ and
$\chi^{\prime\prime}$}

\begin{figure}
\includegraphics[width=7.5cm]{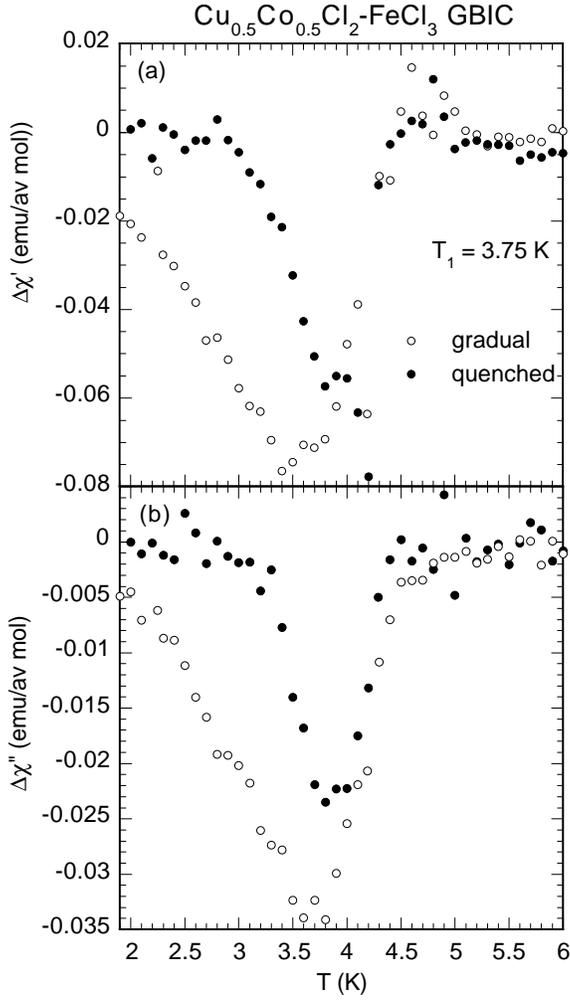}%
\caption{$T$ dependence of (a) $\Delta\chi^{\prime}$ and (b)
$\Delta\chi^{\prime\prime}$: single memory experiments (I and II).  $f =
0.05$ Hz.  $h = 0.1$ Oe.  The measurement (I) (denoted as gradual) was
carried out after the ZFC aging protocol (I): gradual decrease of $T$ from
20 to 3.75 K, isothermal aging at $T_{1} = 3.75$ K for 6.3 hours, and
further gradual decrease from 3.75 to 1.9 K. The measurement (II) (denoted
as quenched) was carried out after the ZFC aging protocol (II): quenching
of the system from 20 to 3.75 K at $H = 0$, isothermal aging at 3.75 K for
10 hours, and further quenching from 3.75 to 1.9 K. Both $\chi^{\prime}$
and $\chi^{\prime\prime}$ were simultaneously measured with increasing $T$
(aging ZFC curves).  $\Delta\chi^{\prime}$ and $\Delta\chi^{\prime\prime}$
are the deviations of the aging ZFC curve from the reference ZFC curve
measured with increasing $T$ after the standard ZFC aging protocol:
quenching from 20 to 1.9 K at $H = 0$.}
\label{fig:eleven}
\end{figure}

The memory effect in the SG system is defined as follows.  When the system
is cooled down to a low temperature below $T_{g}$, a memory of the spin
configurations which is imprinted in the specific cooling sequence, can be
recalled when the system is re-heated at a constant heating rate
\cite{Jonsson1999,Jonason2000}.  In a single memory experiment, the memory
is imprinted at $T_{1}$ for $t_{w1}$ during the ZFC aging protocol.  In a
double memory experiment, the memory is imprinted at $T_{1}$ for $t_{w1}$
and at $T_{2}$ ($< T_{1}$) for $t_{w2}$ during the ZFC aging protocol.  The
dispersion and absorption thus recalled with increasing $T$ are defined as
$\chi^{\prime}_{mem}(\omega,T)$ and $\chi^{\prime\prime}_{mem}(\omega,T)$,
respectively.  For comparison, the dispersion and absorption as references
[$\chi^{\prime}_{ref}(\omega,T)$ and
$\chi^{\prime\prime}_{ref}(\omega,T)]$, are also obtained with increasing
$T$ after the system is quenched from a high temperature above $T_{g}$ to
the lowest temperature.  Such AC susceptibility data are called as the ZFC
reference susceptibilities, where no memory is imprinted.  For clarity we
define the difference between the aging ZFC and reference ZFC
susceptibilities as $\Delta\chi^{\prime}(\omega,T) =
\chi^{\prime}_{mem}(\omega,T)-\chi^{\prime}_{ref}(\omega,T)$ and as
$\Delta\chi^{\prime\prime}(\omega,T) =
\chi^{\prime\prime}_{mem}(\omega,T)-\chi^{\prime\prime}_{ref}(\omega,T)$. 
Figures \ref{fig:eleven}(a) and (b) show the $T$ dependence of
$\Delta\chi^{\prime}$ and $\Delta\chi^{\prime\prime}$ for the single memory
experiment, where $T_{1} = 3.75$ K, $f = 0.05$ Hz, and $h = 0.1$ Oe.  Two
kinds of measurements were carried out, depending on cooling rate during
the ZFC aging protocol: (i) the gradual cooling (gradual decrease of $T$
from 20 to 3.75 K at the cooling rate $5.6 \times 10^{-4}$ K/sec,
isothermal aging at $T_{1} = 3.75$ K for $t_{w1} = 2.27 \times 10^{4}$ sec,
and further gradual decrease from 3.75 to 1.9 K at the rate $2.6 \times
10^{-4}$ K/sec), and (ii) the rapid cooling (quenching of the system from
20 to 3.75 K at $H = 0$ at the rate 0.13 K/sec, isothermal aging at $T_{1}
= 3.75$ K for $t_{w1} = 3.6 \times 10^{4}$ sec, and further quenching from
3.75 to 1.9 K at the rate 0.06 K/sec).  Both $\chi^{\prime}_{mem}$ and
$\chi^{\prime\prime}_{mem}$ were simultaneously measured with increasing
$T$ (aging ZFC curves) from 1.9 to 8 K at the rate $1.7 \times 10^{-4}$
K/sec.  The reference ZFC curves ($\chi^{\prime}_{ref}$ and
$\chi^{\prime\prime}_{ref}$) were measured with increasing $T$ from 1.9 to
8 K at the rate $1.7 \times 10^{-4}$ K/sec after the standard ZFC aging
protocol: quenching from 20 to 1.9 K at $H = 0$ at the rate 0.15 K/sec. 
The results are as follows.  (i) For the case of gradual cooling,
$\Delta\chi^{\prime}$ exhibits negative local minima at 3.4 and 3.75 K,
while for the case of rapid cooling, $\Delta\chi^{\prime}$ exhibits a
negative local minimum at 3.75 K. Above 4.3 K, $\Delta\chi^{\prime}$ is
independent of the detail of the ZFC aging protocol.  (ii) The absorption
$\Delta\chi^{\prime\prime}$ exhibits a negative local minimum at 3.75 K for
both cases of the rapid and gradual cooling.  Above 4.1 K,
$\Delta\chi^{\prime\prime}$ is independent of the detail of the ZFC aging
protocol.  In summary, the equilibration at 3.75 K gives rise to a dip in
$\Delta\chi^{\prime}$ and $\Delta\chi^{\prime\prime}$, suggesting that the
memory of spin configurations which are imprinted during the cooling
process, is recalled during the heating.  The dip of $\Delta\chi^{\prime}$
and $\Delta\chi^{\prime\prime}$ at 3.75 K for the rapid cooling is much
narrower than that for the gradual cooling.  The reason is that in the case
of the gradual cooling the spin configurations at $T$ not equal to 3.75 K
are also imprinted during the cooling process.  Similar but more pronounced
single memory effect has been reported by Jonsson et al.  for Ag (11 at. 
\% Mn) \cite{Jonsson1999}.

\begin{figure}
\includegraphics[width=7.5cm]{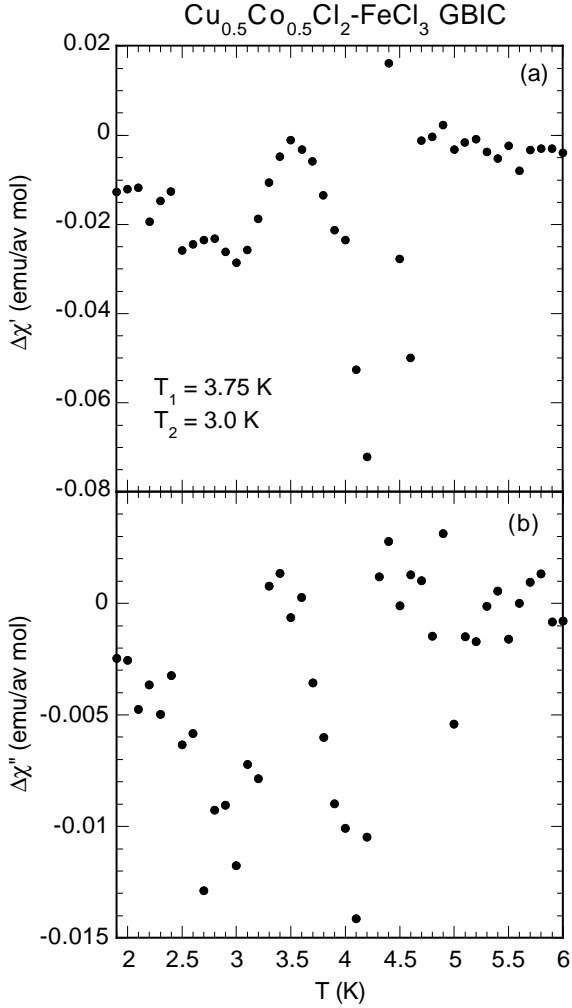}%
\caption{$T$ dependence of (a) $\Delta\chi^{\prime}$ and (b)
$\Delta\chi^{\prime\prime}$: double memory experiment.  $f = 0.05$ Hz.  $h
= 0.1$ Oe.  $H = 0$.  The measurement was carried out after the ZFC aging
protocol: quenching of the system from 20 to 3.75 K at $H = 0$, isothermal
aging at 3.75 K for 10 hours, quenching from 3.75 to 3.0 K, isothermal
aging at 3.0 K for 10 hours, and quenching from 3.0 to 1.9 K. Both
$\chi^{\prime}$ and $\chi^{\prime\prime}$ were simultaneously measured with
increasing $T$ (aging ZFC curves).  $\Delta\chi^{\prime}$ and
$\Delta\chi^{\prime\prime}$ are the deviations of the aging ZFC curve from
the reference ZFC curve measured with increasing $T$ after the standard ZFC
aging protocol: quenching from 20 to 1.9 K at $H = 0$.}
\label{fig:twelve}
\end{figure}

Figures \ref{fig:twelve}(a) and (b) show the $T$ dependence of
$\Delta\chi^{\prime}$ and $\Delta\chi^{\prime\prime}$ for the double memory
experiment, where $T_{1} = 3.75$ K, $T_{2} = 3.0$ K, $f = 0.0$5 Hz, and $h
= 0.1$ Oe.  The measurement was carried out after the ZFC aging protocol
(only in the case of rapid cooling): quenching of the system from 20 to
3.75 K at $H = 0$ at the rate 0.13 K/sec, isothermal aging at $T_{1} =
3.75$ K for $t_{w1} = 3.6 \times 10^{4}$ sec, quenching from 3.75 to 3.0 K
at the rate 0.025 K/sec, isothermal aging at $T_{2} = 3.0$ K for $t_{w2} =
3.6 \times 10^{4}$ sec, and quenching from 3.0 to 1.9 K at the rate 0.04
K/sec.  Both $\chi^{\prime}_{mem}$ and $\chi^{\prime\prime}_{mem}$ were
simultaneously measured with increasing $T$ (aging ZFC curves). 
$\Delta\chi^{\prime}$ and $\Delta\chi^{\prime\prime}$ are the deviations of
the aging ZFC curves from the reference ZFC curves.  The results are as
follows.  Both $\Delta\chi^{\prime}$ and $\Delta\chi^{\prime\prime}$ show a
small dip at $T_{2} = 3.0$ K and a large dip at 4.2 K. There is no dip at
$T_{1} = 3.75$ K. The magnitudes of $\Delta\chi^{\prime}$ and
$\Delta\chi^{\prime\prime}$ for the double memory effect are smaller than
those for the single memory effect.  It seems that the spin configurations
which are imprinted at $T_{1} = 3.75$ K during the cooling process may be
partially reinitialized by the spin configurations imprinted at $T_{2} =
3.0$ K. Similar (but more pronounced) double memory effect has been
reported by Jonsson et al. for Ag (11 at. \% Mn) \cite{Jonsson1999}.  The
condition for the appearance of the dips at $T_{1}$ and $T_{2}$ ($=T_{1}-\Delta
T$) is that the overlap distance $L_{\Delta T}$ given by
Eq.(\ref{eq:nine}) with $T = T_{2}$ is larger than the size
$R_{T_{1}}(t_{w1})$.  The spin configurations imprinted at $T_{1}$ is
partially reinitialized at $T_{2}$, when $\Delta T$ is larger than the
threshold temperature $(\Delta T)_{t}$ given by
\begin{equation} 
(\Delta T)_{t} = (\Upsilon_{T}^{3/2}/T_{1}^{1/2}) 
(t_{w1}/t_{0})^{-\zeta/z(T_{1})}. 
\label{eq:eleven} 
\end{equation} 
When $t_{w1} = 3.6 \times 10^{4}$ sec, $t_{0} = \tau^{*} = 5.29 \times
10^{-6}$ sec, and $\zeta = 0.385$, $(\Delta
T)_{t}$ is estimated as $(\Delta T)_{t} = 0.14 \Upsilon_{T}^{3/2}$.  Using
the value of $\Upsilon_{T}$ ($\approx 1.12 - 1.8$ K) obtained in
Sec.~\ref{resultD}, the value of $(\Delta T)_{t}$ is estimated as $(\Delta
T)_{t} = 0.16 - 0.33$ K, which is smaller than $\Delta T = 0.75$ K.

\subsection{\label{resultF}Rejuvenation effect in 
$\chi^{\prime}$ and $\chi^{\prime\prime}$ under the $T$- and 
$H$-shifts}

\begin{figure}
\includegraphics[width=8.0cm]{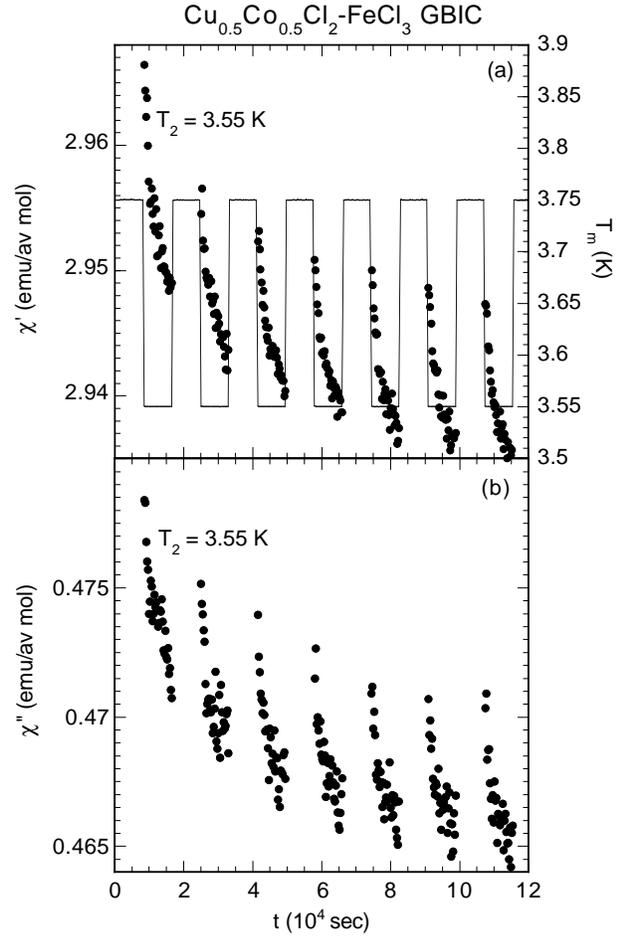}%
\caption{Relaxation of $\chi^{\prime}(\omega,t)$ and
$\chi^{\prime\prime}(\omega,t)$ at $T_{2} = 3.55$ K during a temperature
cycle between $T_{1} = 3.75$ K and $T_{2} = 3.55$ K. The change
of $T$ with $t$ is also shown.  $t = 0$ is a time just after the standard
ZFC aging protocol: quenching of the system from 50 to 3.75 K at $H = 0$. 
The data of $\chi^{\prime\prime}(\omega,t)$ are the same as those presented
in Ref.~\cite{Suzuki2003}.}
\label{fig:thirteen}
\end{figure}

The rejuvenation and memory effects in $\chi^{\prime}$ and
$\chi^{\prime\prime}$ under the $T$-shift are observed in our system. 
Figures \ref{fig:thirteen}(a) and (b) show the $t$ dependence of
$\chi^{\prime}(\omega,t)$ and $\chi^{\prime\prime}(\omega,t)$ at $T_{1} =
3.55$ K under the $T$-shift between $T_{1} = 3.75$ K and $T_{2} = 3.55$ K,
where $f = 0.05$ Hz and $h = 0.1$ Oe.  Here the data at $T_{1} = 3.75$ K
are not shown (see the data of $\chi^{\prime\prime}$ at $T_{1} = 3.75$ K in
the previous paper \cite{Suzuki2003}).  First our system was quenched from
10 K to $T_{1} = 3.75$ K at $H = 0$.  The origin of $t$ ($t = 0$) is a time
when $T$ becomes $T_{1}$.  The relaxation of $\chi^{\prime}$ and
$\chi^{\prime\prime}$ was measured as a function of $t$ during a period
$t_{w1}$ ($\approx 8.2 \times 10^{3}$ sec).  The temperature was then
changed to $T_{2}$ (the negative $T$-shift).  The relaxation of
$\chi^{\prime}$ and $\chi^{\prime\prime}$ was measured as a function of $t$
for a period $t_{w2}$ ($\approx 8.2 \times 10^{3}$ sec) at $T_{2}$.  The
system was again heated back to $T_{1}$ (the positive $T$-shift).  These
processes were repeated subsequently.  Just after every negative $T$-shift,
both $\chi^{\prime}$ and $\chi^{\prime\prime}$ do not lie on the reference
curves of $\chi^{\prime}$ and $\chi^{\prime\prime}$ at $T_{2}$ obtained
when the system is quenched to $T_{2}$ directly from 10 K at $t$ = 0.  The
values of $\chi^{\prime}$ and $\chi^{\prime\prime}$ are larger than the
reference curves, indicating the partial reinitialization (rejuvenation) in
$\chi^{\prime}$ and $\chi^{\prime\prime}$ after the negative $T$-shift. 
Just after every positive $T$-shift, however, both $\chi^{\prime}$ and
$\chi^{\prime\prime}$ lie on to the reference curves of $\chi^{\prime}$ and
$\chi^{\prime\prime}$ at $T_{1}$ obtained when the system is quenched to
$T_{1}$ directly from 10 K at $t = 0$, indicating the memory effect in
$\chi^{\prime}$ and $\chi^{\prime\prime}$ after the positive $T$-shift. 
Note that the reference curve coincides with a curve where the lowest
points for each relaxation in $\chi^{\prime}$ and $\chi^{\prime\prime}$ are
connected as a function of $t$.  The strong rejuvenation effect for the
negative $T$-shift is also predicted from numerical study by Takayama and
Hukushima \cite{Takayama2002} using the MC simulation on the 3D Ising EA SG
model.  Here we note that the threshold temperature differences under the
positive and negative $T$-shift between $T_{1}$ and $T_{2}$, $(\Delta
T)_{+}$ and $(\Delta T)_{-}$, are given by
\begin{eqnarray}
(\Delta T)_{+}
&=&(\Upsilon_{T}^{3/2}/T_{2}^{1/2})(t_{w2}/t_{0})^{-\zeta/z(T_{2})},\nonumber\\
(\Delta T)_{-}
&=&(\Upsilon_{T}^{3/2}/T_{1}^{1/2})(t_{w1}/t_{0})^{-\zeta/z(T_{1})},
\label{eq:twelve} 
\end{eqnarray}
respectively.  When $T_{1} = 3.75$ K, $T_{2} = 3.55$ K, $t_{w1} = t_{w2} =
8.2 \times 10^{3}$ sec, $t_{0} = \tau^{*} = 5.29 \times 10^{-6}$ sec, 
and $\zeta = 0.385$, we have $(\Delta T)_{+} = 0.148
\Upsilon_{T}^{3/2}$ and $(\Delta T)_{-} = 0.163\Upsilon_{T}^{3/2}$:
$(\Delta T)_{+} = 0.18$ K and $(\Delta T)_{-} = 0.19$ K for $\Upsilon_{T} =
1.12$ K. Thus the values of $(\Delta T)_{+}$ and $(\Delta T)_{-}$ are
comparable to the difference $T_{1} - T_{2} = 0.20$ K.

\begin{figure}
\includegraphics[width=7.5cm]{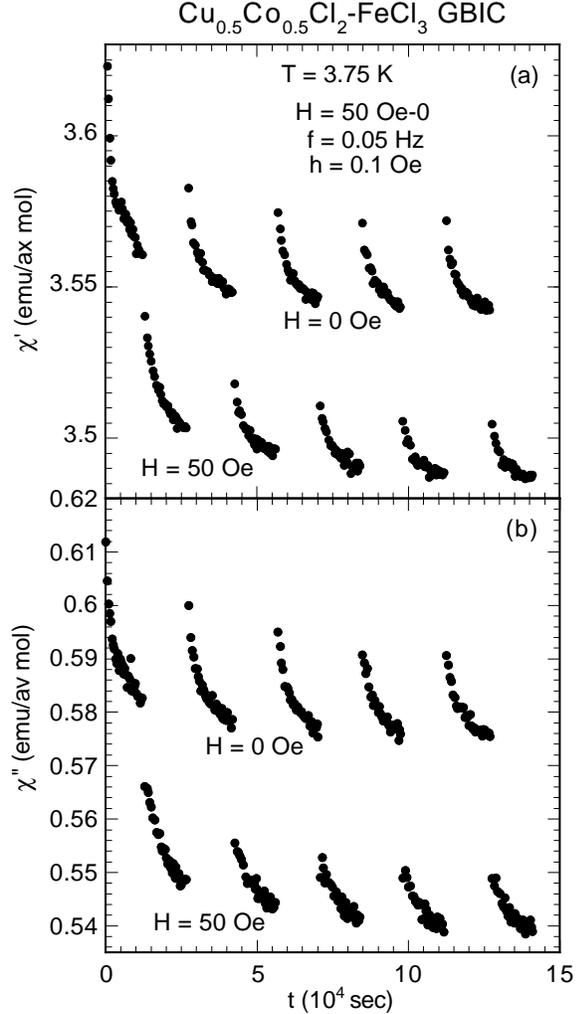}%
\caption{Relaxation of $\chi^{\prime}(\omega,t)$ and
$\chi^{\prime\prime}(\omega,t)$ during a magnetic-field cycle between $H =
0$ and 50 Oe.  $T = 3.75$ K. $t = 0$ is a time just after the standard ZFC
aging protocol: quenching of the system from 50 to 3.75 K at $H = 0$.  The
data of $\chi^{\prime\prime}(\omega,t)$ are the same as those presented in
Ref.~\cite{Suzuki2003}.}
\label{fig:fourteen}
\end{figure}

The rejuvenation effect in $\chi^{\prime}$ and $\chi^{\prime\prime}$ under
the $H$-shift is also observed in our system.  Figures
\ref{fig:fourteen}(a) and (b) show the $t$ dependence of $\chi^{\prime}$
and $\chi^{\prime\prime}$ at $T = 3.75$ K under the $H$-shift, where $f =
0.05$ Hz and $h$ = 0.1 Oe.  After the ZFC aging protocol (quenching from a
high temperature above $T_{g}$ to a temperature $T$ (= 3.75 K) ),
$\chi^{\prime}$ and $\chi^{\prime\prime}$ at $H = 0$ were measured for a
period $t_{w1}$ ($\approx 1.3 \times 10^{4}$ sec), where the origin of $t$
($t = 0$) is a time when $T$ becomes 3.75 K. The field is changed from 0 to
$H$ (= 50 Oe) at $t = t_{w1}$, After this $H$-shift, $\chi^{\prime}$ and
$\chi^{\prime\prime}$ were measured for a period $t_{w2}$ ($\approx 1.3
\times 10^{4}$ sec).  Subsequently, the field was turned off from $H$ to 0
and the measurements were carried out at $H = 0$ for a period $t_{w1}$. 
This process was repeated.  We find that both $\chi^{\prime}$ and
$\chi^{\prime\prime}$ undergo drastic jumps under the $H$-shift (from 0 to
50 Oe) to values higher than the reference curves at $H = 50$ Oe.  They
also undergo drastic jump under the $H$-shift (from 50 to 0 Oe) to values
higher than the reference curves at $H = 0$.  These results suggest that a
partial reinitialization (rejuvenation effect) of $\chi^{\prime}$ and
$\chi^{\prime\prime}$ occurs for the $H$-shifts (0 to 50 Oe and 50 to 0
Oe).  Note that the point at $T = 3.75$ K and $H \approx 50$ Oe is located
on the de Almeida-Thouless (AT) \cite{Almeida1978} line in the ($H$,$T$)
phase diagram \cite{Suzuki2003}.  On this AT line, the $T$-derivative
d$\delta$/d$T$ shows a local minimum, where $\delta =
\chi_{FC}-\chi_{ZFC}$.

When the overlap distance $L_{H}$ given by Eq.(\ref{eq:seven}) is smaller
than the size $R_{T}(t_{w})$, the spin configuration imprinted at $H = 0$
is partially reinitialized at $H$ ($> H_{t}$), where the threshold magnetic
field is defined as
\begin{equation} 
H_{t}=\Upsilon_{H}(t_{w}/t_{0})^{-\delta/z(T)}. 
\label{eq:thirteen} 
\end{equation} 
When $t_{w} = 3.6 \times 10^{4}$ sec, $t_{0} = \tau^{*} = 5.29 \times
10^{-6}$ sec, $\delta = 1.37$, and $T = 3.75$ K,
$H_{t}$ is estimated as $H_{t} = 8.67 \times 10^{-3} \Upsilon_{H}$.  Using
the value of $\Upsilon_{H}$ (= 2 - 3 kOe) obtained in Sec.~\ref{resultC},
the value of $H_{t}$ is estimated as 17 - 25 Oe, which is lower than 50 Oe.

\section{\label{dis}Discussion and conclusion}
The non-equilibrium nature of the spin dynamics in 3D Ising SG
Cu$_{0.5}$Co$_{0.5}$Cl$_{2}$-FeCl$_{3}$ GBIC has been studied from the $t$
dependence of $\chi_{ZFC}$, $\chi^{\prime}$, and $\chi^{\prime\prime}$
after specific ZFC protocols including the $T$- shift ($\Delta T$) and
$H$-shifts.  The relaxation rate $S(t)$ shows a peak at $t_{cr}$ ($\approx
t_{w}$), corresponding to a crossover from quasi-equilibrium dynamics to
non-equilibrium dynamics.  The value of $t_{cr}$ strongly depends on the
wait time $t_{w}$, $T$, $H$, and $\Delta T$.  The rejuvenation effects are
observed in $\chi^{\prime}$ and $\chi^{\prime\prime}$ under the negative
$T$-shift and both the positive and negative $H$-shifts.  The spin
configurations imprinted under the ZFC aging protocols are recalled on
heating the system.

We find that the observed change of $t_{cr}$ under the $T$-and 
$H$-shifts is well explained in terms of the scaling relations where the
overlap lengths $L_{\Delta T}$ and $L_{H}$ play a significant role.  Under
the $T$-shift from $T = T_{i}$ to $T_{f}$ ($=T_{i} +\Delta T$), the size of
domains are unaffected for sufficiently small $\vert\Delta T\vert$, where
$L_{\Delta T}$ is larger than $R_{Ti}(t_{w})$.  Then the
relaxation rate $S(t)$ shows a peak at $t_{cr}$ where 
$L_{T_{f}}(t)$ is equal to $R_{T_{i}}(t_{w})$.  In contrast, the size
of domains are affected for sufficiently large $\vert\Delta T\vert$.  
The overlap
length $L_{\Delta T}$ becomes lower than $R_{T_{i}}(t_{w})$.  Then $S(t)$
has a peak at $t_{cr}$, where $L_{T_{f}}(t)$ is
equal to $L_{\Delta T}$.  Under the $H$-shift, the size of domains are
unaffected during the $H$ shift for sufficiently small $H$, where the
overlap length $L_{H}$ is larger than $R_{T}(t_{w})$.  The value of
$t_{cr}$ is dependent on $t_{w}$.  In contrast, the size of domains are
affected during the $H$ shift for sufficiently large $H$, where $L_{H}$
becomes lower than $R_{T}(t_{w})$.  Then $S(t)$ has a peak at $t_{cr}$,
where $L_{T}(t) \approx L_{H}$.

We discuss the scaling relation of the $T$ dependent relaxation rate
$S(T,t)$.  As pointed out in Sec.~\ref{resultA} $\chi_{ZFC}(T,t)$ below
$T_{g}$ may be described by a scaling function
\begin{equation} 
\chi_{ZFC}(T,t) = G(x),
\label{eq:fourteen} 
\end{equation} 
with $x = L_{T}(t)/R_{T}(t_{w})$.  From the definition, $S(T,t)$ is derived
as
\begin{equation} 
S(T,t) = {\rm d}\chi_{ZFC}(T,t)/{\rm d}\ln t = (1/z(T)) H(x),
\label{eq:fifteen} 
\end{equation} 
with $H(x) = x($d$G(x)/$d$x)$.  It follows that $S(t)$ is described by a
scaling function $H(x)$ except for the factor $1/z(T)$.  When the scaling
function $H(x)$ has a peak at $x = a$ (constant), then $S(t)$ has a peak at
$t = t_{cr}$, where $x = L_{T}(t)/R_{T}(t_{w}) = a$.  Then $t_{cr}$ is
simply described as
\begin{equation} 
\ln(t_{cr}/t_{w}) = (T_{g}/bT) \ln a.
\label{eq:sixteen} 
\end{equation} 
When $a$ is larger than 1, $\ln(t_{cr}/t_{w})$ increases with decreasing
$T$.  In fact, the least squares fit of the data of $\ln(t_{cr}/t_{w})$ vs
$t$ with $t_{w} = 2.0 \times 10^{3}$ sec in Fig.~\ref{fig:three}(a) ($H =
1$ Oe) to Eq.(\ref{eq:sixteen}) yields to $(T_{g}/b) \ln a = 37.1 \pm
1.9$.  Since $T_{g} = 3.92$ K and $b = 0.16$, we have $a = 4.5 \pm 0.4$. 
It is predicted from Eq.(\ref{eq:fifteen}) that $S_{max}$ increases with
increasing $T$ for $T < T_{g}$ since $S_{max}$ is linearly dependent on the
factor $1/z(T)$ ($= bT/T_{g}$).  Experimentally, as shown in
Fig.~\ref{fig:three}(b) with $t_{w} = 2.0 \times 10^{3}$ sec, the peak
height $S_{max}$ strongly depends on $T$.  The peak height $S_{max}$
exhibits a broad peak around 3.6 - 3.8 K, just below $T_{g}$. 
The linear increase of $S_{max}$ with increasing $T$ below $T_{g}$ is
considered to be due to the factor $1/z(T)$.  The decrease of $S_{max}$
with increasing $T$ above $T_{g}$, however, cannot be explained in terms of
the above model because the scaling relation is valid only for $T < T_{g}$. 
In Sec.~\ref{resultA} we show that $S(t)/S_{max}$ at $T = 3.75$ K and $H =
5$ Oe obeys a $t/t_{w}$-scaling law for long $t_{w}$ ($5.0 \times
10^{3}\leq t \leq 1.5 \times 10^{4}$ sec) and $0 \leq t \leq 6.0 \times
10^{4}$ sec (see Fig.~\ref{fig:four}(c)): $S(t)/S_{max}$ is well described
by a scaling function $F(t/t_{w})$ which has a peak at $t_{cr}/t_{w}
\approx 0.68$.  Using Eq.(\ref{eq:sixteen}), the value of $a$ is estimated
as $a = 1.11$, indicating that the constant $a$ at $H = 5$ Oe is different
from that at $H = 1$ Oe.

Finally we consider the cause of the complicated behavior of $S(t)$ at $T =
3.75$ K for $0 \leq t_{w} \leq 750$ sec.  One of the cause is the way how $T$
approaches 3.75 K during the ZFC aging protocol.  The temperature drops
rapidly to $\approx 3.50$ K and slowly approaches $3.75 \pm 0.01$ K from
the below (usually) within 230 sec.  The experiment of $S(t)$ under the $T$
shift (see Sec.~\ref{resultD}) suggests that the initial undercool is not
so important because of the temperature difference larger than the 
threshould 
temperature difference $(\Delta T)_{t}$ \cite{Zotev2003}.  However, the
subsequent approach of $T$ to 3.75 K for a wait time $t_{w0}$ ($\leq 230$
sec) may play a significant role in the aging behavior after $t = $0 at
$T_{m} = 3.75$ K. The threshold temperature difference $(\Delta T)_{t}$ is
estimated as
\begin{equation} 
(\Delta T)_{t}=(\Upsilon_{T}^{3/2}/[(T_{m}-(\Delta T)_{t})^{1/2}]) 
(t_{w0}/t_{0})^{-\zeta/z(T_{m}-(\Delta T)_{t})}, 
\label{eq:seventeen} 
\end{equation} 
from the condition that $L_{\Delta T} = R_{T-\Delta T}(t_{w0})$.  When
$t_{0} = \tau^{*} = 5.29 \times 10^{-6}$ sec, $\zeta = 0.385$, and $t_{w0}
= 230$ sec, $(\Delta T)_{t}$ is estimated as $(\Delta T)_{t} = 0.24$ K for
$\Upsilon_{T} = 1.12$ K and 0.56 K for $\Upsilon_{T} = 1.80$ K. This
suggests that the spin configuration at $T_{m} = 3.75$ K after $t = 0$ is
affected by that imprinted at $T_{m}-\Delta T$ at the wait time $t_{w0}$. 
Note that the value of $t_{cr}$ at $T_{m}-\Delta T$ is larger than that at
$T_{m}$ for the same $t_{w}$ (see Fig.~\ref{fig:three}(a)).  
The appearance of two peaks in $S$ vs $t$ at
$0 \leq t_{w} \leq 200$ sec may be associated with two domains generated
for the wait time $t_{w0}$ at $T_{m}-\Delta T$ and for $t_{w}$ at $T_{m}$.

In conclusion, we have undertaken an extensive study on the $t$ dependence
of the relaxation rate $S(t)$ mainly below $T_{g}$ under the various
conditions including the $T$- and $H$-shifts.  The $t$ dependence of
$S$ is well explained in terms of the scaling relation.  The peak of
$S(t)$ occurs when the mean SG domain size $L_{T}(t)$ coincides with
$R_{T}(t_{w})$, where $t_{w}$ is the wait time.  Our results indicate that
the aging, memory and rejuvenation phenomena observed in our system are
very similar to those in conventional spin glass systems.

\begin{acknowledgement}
We would like to thank H. Suematsu for providing us with single crystal
kish graphite and T. Shima and B. Olson for their assistance in sample
preparation and x-ray characterization.  Early work, in particular for the
sample preparation, was supported by NSF DMR 9201656.
\end{acknowledgement}
%
%

\end{document}